# Non-universal Impact of Cholesterol on Ionic Liquid-Membrane Interactions

J. Gupta[1,2], V. K. Sharma[1,2*], P. Hitaishi[3$], A. K. Jha[4], J. B. Mitra[5], H. Srinivasan[1], S. Kumar[1,2], A. Kumar[4], S. K. Ghosh[3], S. Mitra[1,2]

[1]Solid State Physics *Division, Bhabha Atomic Research Centre, Mumbai, 40008, India*
[2]*Homi Bhabha National Institute, Mumbai, 400094, India*
[3]*Department of Physics, School of Natural Sciences, Shiv Nadar Institution of Eminence, NH 91, Tehsil Dadri, G. B. Nagar, Uttar Pradesh 201314, India*
[$]*Institute for Experimental and Applied Physics, Christian-Albrechts-University of Kiel, 24118 Kiel, Germany* (present address)
[4] *Department of Biosciences and Bioengineering, Indian Institute of Technology Bombay, Mumbai, 400 076, India*
[5]*Radiopharmaceuticals Division, Bhabha Atomic Research Centre, Mumbai 400085, India*

*Abstract*

Understanding the role of cholesterol in ionic liquid (IL)-membrane interactions is essential for advancing biomedical applications of ILs, including the development of innovative antimicrobial agents. In this study, we explore the intricate and multifaceted role of cholesterol in modulating IL-membrane interactions, employing a comprehensive suite of biophysical techniques. We systematically examine how IL alkyl chain length and membrane physical state influence the impact of cholesterol on IL-lipid membrane interaction. The incorporation of ILs is shown to increase the area per lipid in both pristine dipalmitoylphosphatidylcholine (DPPC) and DPPC-cholesterol membranes. Cholesterol modulates the impact of ILs on lipid conformation, membrane viscoelasticity, and phase behavior. Small-angle neutron scattering and dynamic light scattering measurements reveal that cholesterol mitigates IL-induced structural perturbations in vesicles. Our isothermal titration calorimetry measurements reveal that the presence of cholesterol significantly weakens the binding of ILs to membranes. Intriguingly, despite this reduced binding affinity, cholesterol-containing membranes demonstrate enhanced permeabilization. This counterintuitive effect is attributed to cholesterol's ordering of lipid membranes, which increases susceptibility to stress and defects. Our results underscore the complex and non-universal interplay between lipid composition, IL alkyl chain length, and membrane phase state. These insights provide a deeper understanding of cholesterol's role in IL-membrane interactions, paving the way for the design of advanced applications of ILs in antimicrobial therapy and drug delivery.

* Email: sharmavk@barc.gov.in;   Phone +91-22-25594604

# 1. INTRODUCTION

Ionic liquids (ILs) constitute a unique category of organic salts in which ions have limited coordination, leading to a melting point that falls below 100 °C[1-3]. ILs possess distinctive properties like low volatility, high thermal stability and ionic conductivity, and have tunable chemical composition. These characteristics make them valuable in various applications including pharmaceutical industries[4]. Numerous research groups have highlighted the antimicrobial potential of ILs[3, 5]. Nonetheless, a comprehensive understanding of the fundamental mechanisms governing the antimicrobial activities of ILs is still elusive. A profound understanding of these mechanisms is essential to streamline and enhance the development of applications in the field of bio-nanomedicine involving IL. There are several studies suggesting that the actions of ILs are associated with their perturbing effects on cellular membranes[5-7]. ILs may affect the cell membrane via various methods such as (i) adsorption of ILs onto the cell membrane, (ii) electrostatic interactions between ILs and the membrane, and (iii) the penetration of ILs into the membrane. Consequently, the presence of ILs alters the characteristics of cell membrane, impacting its fluidity, viscoelasticity[3]. Alterations in cell membrane fluidity, in turn, affect the diffusion rate and stability of proteins. ILs may also alter the membrane permeability by creating pores that result in irreversible damage[8]. The lipophilicity of ILs plays a pivotal role in their antimicrobial activity and primarily depends on the type of cation and hydrophobicity of the molecules. This versatility makes ILs promising candidates as foundational components for the development of novel antimicrobial agents. However, for these ILs to function therapeutically without posing a risk of harm to humans, it's crucial that they exhibit selectivity, specifically targeting bacterial membranes. Bacterial and mammalian cell membranes exhibit contrast in their chemical compositions. A key difference is the abundance of cholesterol (up to 50 mol%) in eukaryotic cell membranes, absent in their bacterial counterparts[9]. This abundance significantly influences the thermodynamical, physical, and biological properties of cell membrane[10-15].

Cholesterol features a small polar hydrophilic hydroxyl group and a relatively larger hydrophobic steroid ring, which lacks the ability to self-assemble into distinct structures. However, its polar hydroxyl group aligns with the ester carbonyl groups of the phospholipids, with which it is perpendicularly embedded. The "umbrella model" suggests that the lipid head group shields the hydrophobic steroid rings of cholesterol from unfavourable interactions with water[16]. It involves the straightening and changing in the orientation of lipid hydrophobic alkyl chains, leading to the cholesterol "condensing effect". While cholesterol serves various functions within cells, it plays a pivotal role in modulating the physical properties of the plasma membrane such as mechanical strength (bending and compressibility modules), fluidity, and so on. Effects of cholesterol on the membrane dynamics have been investigated using quasielastic neutron scattering (QENS) and shown that the presence of cholesterol restricts lateral diffusion of lipids in the fluid phase ($L_\alpha$)[17]. This restriction can be attributed to cholesterol's ability to promote tight packing of lipid molecules, creating hindrance to their lateral mobility. Cholesterol exhibits intriguing properties, it promotes ordering in the fluid phase while exerting an opposite influence in the gel phase ($L_\beta$)[18]. Consequently, the presence of cholesterol in the membrane serves as a regulatory factor, preventing them from exclusively adopting either the gel phase or the fluid phase. The distribution of cholesterol in these membranes is believed to be inhomogeneous and leads to the formation of distinct cholesterol-rich and cholesterol-poor domains, signifying phase segregation. Such cholesterol-rich domains are commonly referred to as lipid rafts, which play a pivotal role in diverse cellular activities such as signaling, membrane fusion, and membrane trafficking[19]. It is evident that cholesterol influences diverse biophysical properties of the membrane, and its impact on the lipid bilayer may extend to affect interactions with IL-lipid bilayers. Nevertheless, this aspect has received minimal attention. There are only a few reports[12-13, 20-21] that explore the investigation of the role of cholesterol in IL-membrane interactions. The focus of these investigations has centered on lipids like phosphatidylcholine (PC) and sphingomyelin (SM). These lipids, along with cholesterol, constitute the primary components of plasma membranes in numerous mammalian cells. For instance, Russo et al.[20] investigated the interactions between cholesterol-richPC-based lipid vesicles and trioctylmethylphosphonium acetate ILs. Their findings revealed that lipid bilayers enriched with cholesterol exhibit reduced susceptibility to membrane-disrupting agents compared to membranes lacking any sterols. Recently, Hao et al.[13] utilized egg SM as a model membrane system, demonstrating that cholesterol serves as a protective factor against the destructive impact of ILs. Their findings indicate that in the absence of cholesterol, a minimal IL:SM molar ratio of 0.01:1 can compromise the integrity of the bilayer structure. In stark contrast, the inclusion of cholesterol in lipid bilayers enables the liquid-ordered phase to withstand the deleterious effects of the ILs even up to much higher concentrations of ILs.

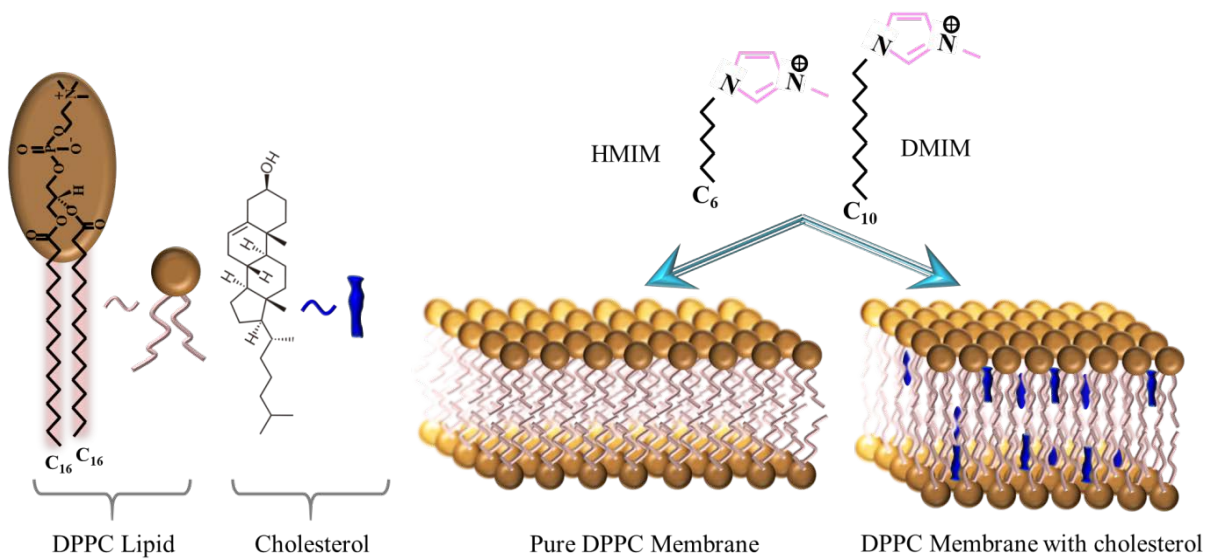

**FIGURE 1:** Molecular structures of zwitterionic dipalmitoylphosphatidylcholine (DPPC) lipid, cholesterol, and cations of HMIM[Br] and DMIM[Br] ILs. A schematic representation illustrating the interactions between ILs and membranes, both in the absence and presence of cholesterol.

We opted dipalmitoylphosphatidylcholine (DPPC), a saturated PC, whose transition temperatures fall in an experimentally convenient temperature range. This choice allows us to investigate the impact of the membrane's physical state on its interactions with ILs. Here, we report the role of cholesterol in the interaction of imidazolium based ILs with DPPC membranes. We have selected two imidazolium based ILs, 1-decyl-3-methylimidazolium bromide (DMIM[Br] or $C_{10}$MIM[Br]) and 1-hexyl-3-methylimidazolium bromide (HMIM[Br] or $C_6$MIM[Br]), which share the same head group but differ in alkyl chain lengths to investigate the role of alkyl chain lengths. Two model membrane systems, DPPC and DPPC supplemented with 20 mol % cholesterol have been chosen. The molecular configurations of zwitterionic DPPC, cholesterol and the respective ILs, along with a schematic of the DPPC membrane in the presence and absence of cholesterol, are shown in Fig. 1. We have employed a combination of various biophysical methods such as pressure-area isotherm, in-pane viscoelasticity, differential scanning calorimetry (DSC), Fourier transform infrared (FTIR) spectroscopy, dye leakage, isothermal titration calorimetry (ITC), dynamic light scattering (DLS), small angle neutron scattering (SANS), on model membrane systems with and without ILs. Analyzing the outcomes derived from the addition of ILs in both cholesterol-free and cholesterol-supplemented bilayers showed non-universal and complex influence of cholesterol on IL-lipid interactions.

## 2. MATERIALS AND METHODS

### 2.1. Materials:

DPPC (>97 %) in powder form was purchased from Avanti Polar Lipids (Alabaster, AL) and used as received without further purification. Cholesterol, HPLC spectroscopy grade chloroform (purity > 99.9%), and $D_2O$ (99.9% purity) were obtained from Sigma Aldrich. The ILs, HMIM[Br] and DMIM[Br], were procured from Tokyo Chemical Industries Co. Ltd. The fluorescent dye calcein and the nonionic surfactant Triton X-100 were acquired from Sisco Research Laboratories Pvt. Ltd, Maharashtra, India. De-ionized (DI) (Milli-Q, Millipore) water with resistivity of ~ 18 MΩ-cm and pH ~ 7.5 was used.

### 2.2. Sample preparations:
Two different kinds of model systems, monolayer and unilameller vesicles (ULVs) have been used. DPPC, cholesterol and ILs were individually dissolved in chloroform to prepare a stock solution with a concentration of 0.5 mg/ml. Later, the solutions of IL and DPPC were combined to achieve a specific mole percentage (mol %). The amount of IL in the lipid/ cholesterol/IL mixture was quantified as,

$$\text{mol\% of IL} = \frac{[IL]}{[IL] + [Lipid] + [Cholesterol]} \times 100\% \qquad (1)$$

where, $[IL]$, $[Lipid]$ and $[Cholesterol]$ are the molar concentrations of IL, lipid and cholesterol, respectively. This mixture is used to prepare a monolayer at air-water interface. Pressure-area isotherm and in-plane viscoelasticity measurements were carried out on the monolayer. ULVs were prepared via the extrusion method as described elsewhere[22-24]. In brief, DPPC lipid powder, and DPPC lipid powder with 20 mol % cholesterol were dissolved in chloroform in two separate vials. Chloroform was gently evaporated with nitrogen gas, and vacuum-dried to form dry lipid films. These films were then hydrated with aqueous solution at 330 K, vortex-mixed, and extruded more than 21 times through a mini-extruder with a 100 nm porous polycarbonate membrane at 330 K. To introduce HMIM[Br] and DMIM[Br] ILs into the membranes, the respective ILs were added to the vesicle solution. DSC, FTIR, ITC, DLS, and SANS measurements were carried out on ULVs samples.

### 2.3. Surface pressure (π) – area (A) isotherm
Interaction of an IL with lipid DPPC and DPPC/cholesterol monolayers have been quantified using π-A isotherms and in-plane dilation rheology. A Langmuir-Blodgett (LB) trough of size 36.4 × 7.5 × 0.4 cm$^3$ (KSV NIMA, Biolin Scientific) with two symmetric Delrin barriers and a platinum Wilhelmy balance was used to record the isotherms of monolayers. The temperature of water filled in the trough was maintained to 300 K by circulating water through the bottom base of the trough using a water bath (Equibath, India). This temperature is below the main phase transition temperature ($T_m$) of the DPPC, therefore, the lipids in the monolayer at air-water interface remain in the gel phase. To generate an isotherm, the surface pressure was monitored as a function of the area per molecule (APM). A specific volume of lipid, mixed lipid/IL or lipid/cholesterol/IL solution was spread on the water surface of the trough using a glass Hamilton micro syringe, followed by 20 minutes waiting time to allow complete evaporation of chloroform. The monolayer was compressed at a constant rate of 4 mm/min until it reached the collapse pressure, or barriers reached their extreme limit.

### 2.3. Dilation rheology of lipid monolayer

The in-plane rheology measurements were performed using the same LB trough where a monolayer was compressed at a slow rate of 2 mm/min to a target surface pressure and then the two symmetrical barriers were oscillated at a specific frequency. This oscillation of barriers generated a time dependent, sinusoidal strain in surface area. As a response, the time-dependent variation in surface pressure was recorded as the stress. The measurements were performed by applying a strain magnitude of 1% in relative to the surface area of monolayer. The barriers were oscillated over a frequency range of 70 to 370 mHz.

The following equations were used to estimate the dynamic viscoelasticity of the monolayer[25-27],

$$a(t) = A(1 + a_0 \sin(\omega t + \phi_a)) \quad (2)$$

$$\pi(t) = P + \pi_0 \sin(\omega t + \phi_\pi) \quad (3)$$

where $A$, $a_0$, and $a(t)$ are the initial surface-area, amplitude of applied strain (1% of $A$) and time dependent change in the area, respectively. $\omega$ is the angular frequency of oscillation, $\pi_0$ is the measured stress amplitude and $P$ is the initial surface pressure. $\pi(t)$ is the time-dependent response in monolayer surface pressure. Because of the viscoelastic nature of the film, there is a phase lag between stress and strain, which is given by $\phi = \phi_\pi - \phi_a$. The two viscoelastic parameters of the monolayer, namely, the storage modulus ($E'$) and loss modulus ($E''$) are expressed in term of this measured phase lag by the equations[8, 26, 28-29]

$$E' = \frac{\pi_0}{a_0} \cos\phi$$

$$E'' = \frac{\pi_0}{a_0} \sin\phi \quad (4)$$

while the storage modulus ($E'$) is related to the elastic nature of the monolayer, the loss modulus ($E''$) is connected to the viscous nature of the film.

**2.4. Dye leakage assay**

To investigate the role of cholesterol on the membrane's permeability, dye leakage assays were carried out on DPPC membranes with cholesterol at varying concentrations of ILs. In this method, DPPC with 20 mol% cholesterol films were hydrated with the green fluorescent dye calcein at a concentration of 70 mM, prepared in 1 M NaOH and 1X PBS buffer (pH 7.4). The resulting samples were vortex-mixed, subjected to three freeze-thaw cycles, and then extruded as discussed above. To separate the non-trapped dye molecules (green in color) from dye- trapped ULVs (dark brown in color), the extruded stock solution passed through a Sephadex G-25 column pre-equilibrated with 1X PBS. This process, known as size exclusion chromatography. To remove the remaining dye residues, solution was poured into a dialysis tube composed of cellulose membrane (cut off = 16 kDa) and placed overnight in 1X PBS buffer. The selective pore size of the dialysis cellulose membrane allowed non-trapped dye to escape into the buffer, leaving the vesicles trapped dye molecules inside. Dye leakage experiments were conducted in a 96-well plate format, utilizing calcein-loaded ULVs at a final lipid concentration of ~270 µM. Various concentrations of HMIM[Br] and DMIM[Br] were used, and volumes were adjusted with 1X PBS buffer to 100 µL. Percent dye release was calculated by normalizing the observed fluorescence intensity at a given time.

$$\text{Dye release (\%)} = \frac{F_s(t) - F_0(t)}{F_{100}(t) - F_0(t)} \times 100 \tag{5}$$

Where $F_s(t)$ is the fluorescence intensity for the given sample at time $t$. Untreated ULVs were considered to have 0% release or no release of the dye ($F_0$), while Triton X-100 (1% v/v) treated ULVs served as a positive control with 100% dye release ($F_{100}$). Dye release over time (up to 30 minutes) was monitored by tracking changes in fluorescence intensity, with a dead time of 30 seconds. The interaction of the ILs with the lipid membrane was observed using a plate reader (Polarstar omega, BMG Labtech, Offenburg, Germany) at 320 K. Calcein emission was measured at 520 nm, with an excitation wavelength of 485 nm. To ensure reproducibility, measurements were performed on three independent sets of samples.

## 2.5. FTIR measurement

Infrared spectroscopic studied were performed on DPPC with and without cholesterol ULVs prepared in $D_2O$ at varying concentrations of HMIM[Br] and DMIM[Br] at different temperatures. These investigations employed a Bruker IFS125 Fourier transform spectrometer equipped with a liquid nitrogen-cooled MCT detector, KBr beam splitter, a globar source and heatable sample stage. The liquid cell sample chamber utilized two $CaF_2$ windows with a 10 mm diameter and a ~60 μm path length. About 40 μL of ULVs sample was filled into the sample chamber for measurements. The sample chamber was purged with dry nitrogen gas. All the measurements were carried out in the transmission mode. For analysis, the data were converted to absorbance mode using the relation $Ab = -ln(Tr)$, where *Ab* is the absorbance, *Tr* is the transmittance, and *ln* is the natural logarithm. Data were collected in the mid-infrared region (900-4000 $cm^{-1}$) at temperature range of 303-328 K in 1 K/min increments. Background correction included data from an empty cell and $D_2O$. Spectra were fitted with a Lorentzian lineshape, and each was co-added with 120-scan, maintaining minimal resolution of 2 $cm^{-1}$.

## 2.6. DLS measurement

The Zetasizer Nano ZS system (Malvern Instruments, U.K.), equipped with a 633 nm He-Ne laser, was used to perform DLS measurements on DPPC with 20 mol% cholesterol ULVs prepared in $D_2O$ at various concentrations of HMIM[Br] and DMIM[Br] ILs. Measurements were carried out at a scattering angle of 173°. Prior to measurement, samples were diluted to a concentration ~ 2 mM and passed through a 0.45 μm MILLEX-HV syringe filter to ensure that no dust particles were present. The filtered samples were then placed in disposable sizing cuvettes for measurement. DLS measurements were taken at 300 K and 330 K. Prior to each measurement, the samples were thermally equilibrated for 5 minutes.

## 2.7. SANS measurement

SANS measurements were carried out on ULVs of 30mM DPPC with 20 mol % cholesterol prepared in $D_2O$ at varying concentrations of HMIM[Br] and DMIM[Br]. For each sample, measurements were carried out at 300 K and 330 K. The measurements were performed at SANS facility at the Dhruva Reactor, Bhabha Atomic Research Centre, Mumbai, India[30]. A monochromatic beam of neutrons (wavelength $\lambda \sim 5.2$ Å, $\Delta\lambda/\lambda \sim 15\%$) was directed onto the samples, and the scattered neutrons were captured using a $^3$He detector. The instrument covers a scattering vector ($Q = (4\pi \, Sin \, \theta)/\lambda$; $2\theta$ is the scattering angle) range of 0.01~0.3 $Å^{-1}$.

Subsequently, the data were corrected for transmission, empty cell contribution, backgrounds and normalized to an absolute scale using standard protocols.

## 2.8. ITC measurements

As previously shown[5], longer chain IL, DMIM[Br] interacts more strongly with DPPC membrane compared to HMIM[Br]. Here, to investigate role of membrane composition (presence of cholesterol) on binding characteristics of DMIM[Br] with DPPC membrane, ITC measurements were carried out. Experiments were conducted using the MicroCal iTC200 system from Malvern Instruments. The calorimeter cell (200 μL) was loaded with 0.5 mM DPPC with 20 mol % cholesterol ULVs prepared in PBS. A 40 µL syringe of the calorimeter was filled with 6 mM DMIM[Br] for DPPC with cholesterol. The reference cell was filled with DI water. DMIM[Br] was injected into DPPC with cholesterol ULVs under continuous stirring at 1000 rpm until the signal saturation was achieved. A total of 23 injections were performed in sequential aliquots of $1 \times 0.4$ μL, $19 \times 1.5$ μL, and $3 \times 3$ μL. The first injection, known to create equilibration artifacts, was excluded from the analysis. In ITC, molecular interactions induce changes in enthalpy at each injection, observed as peaks in a plot of heat flow over time. Integration of these peaks provides the enthalpy change in Kcal/mol relative to the molar ratio of DMIM[Br]. Control experiments injecting the DMIM[Br] into buffer under similar conditions were conducted to subtract the contribution of IL-solvent interactions from the total heat change per injection.

## 2.9. DSC measurements

DSC experiments were carried out on 20 mM DPPC and DPPC with cholesterol ULVs prepared in MQ water with and without 50 mol % HMIM[Br] and DMIM[Br] using Malvern MicroCal PEAQ DSC. The thermograms were recorded in the temperature range of 290 K to 330 K in the heating cycle with a fixed scan rate of 1 K/min.

## 3. RESULTS AND DISCCUSION

### 3.1. Impact of cholesterol on the IL-lipid interaction in a monolayer

A Langmuir-Blodgett trough is widely used to study the interaction of various membrane active molecules to a single molecular layer of amphiphiles formed at the air-water interface[25, 31-34]. Researchers have used this methodology recently to understand the interaction of ILs with lipid monolayer[7, 23, 25, 35]. In the present study, the monolayer is formed by homogenous mixture of DPPC and cholesterol and two ILs at different concentrations, which was spread on the water surface[12, 25, 35-38]. Isotherms are generated by recording the variation in surface pressure on compression of the monolayer by bringing the barriers close to each other and are shown in Fig. 2. Here, in all cases, the APM is calculated by considering the number of DPPC and cholesterol molecules spread at the interface. Therefore, the shift of the isotherm to a higher APM is the signature of the interaction of the ILs with lipid monolayer. Note that the concentration of ILs in DPPC/IL and DPPC/cholesterol/IL mixed system is increased in such a way that the number of lipid molecules spread on the interface remains constant. Initially, barriers are at the extreme position with the molecules being far apart from each other. Due to negligible intermolecular interaction, there is no measurable surface pressure. This region is referred to as gaseous (G) phase. Compression of barriers brings the molecules close enough that they start interacting, showing an increase in surface pressure. This region is known as liquid expanded (LE) phase. The area at which the pressure just starts to rise is known as lift-off area ($A_L$). The data clearly indicate that the $A_L$ in the DPPC monolayer increases upon incorporation of ILs. Specifically, the presence of 10 and 25 mol% of the HMIM[Br] leads to a ~ 3 % and 6 % increase in $A_L$, observed in Fig. 2 (a). In contrast, the effect is significantly more pronounced with DMIM[Br] where $A_L$ increases by 13 % and 20 % (Fig. 2 (b)) at the same IL concentrations. These results strongly suggest that DMIM[Br] perturbs the lipid monolayer to a much greater extent than HMIM[Br]. Upon further compression, the monolayer transitions from a LE to a liquid-condensed (LC) phase, characterized by more tightly packed lipid molecules. For pure DPPC, a distinct plateau region is observed between the LE and LC phases, indicating the coexistence of both states. However, in the presence of either IL, this plateau shifts to higher surface pressures and becomes less defined, reflecting a disruption in phase coexistence. The extension of the LE phase to higher surface pressures highlights the disordering influence of both ILs on the membrane structure, with DMIM[Br] exerting a notably stronger effect[25].

The addition of 20 mol% cholesterol in the DPPC membrane reduces the area occupied by the molecules and leads to the formation of DPPC/ cholesterol mixed lipid monolayer with a compact arrangement of molecules. The addition of ILs in DPPC/ cholesterol monolayer exhibits a similar isotherm trend to that observed with the pure DPPC monolayer.

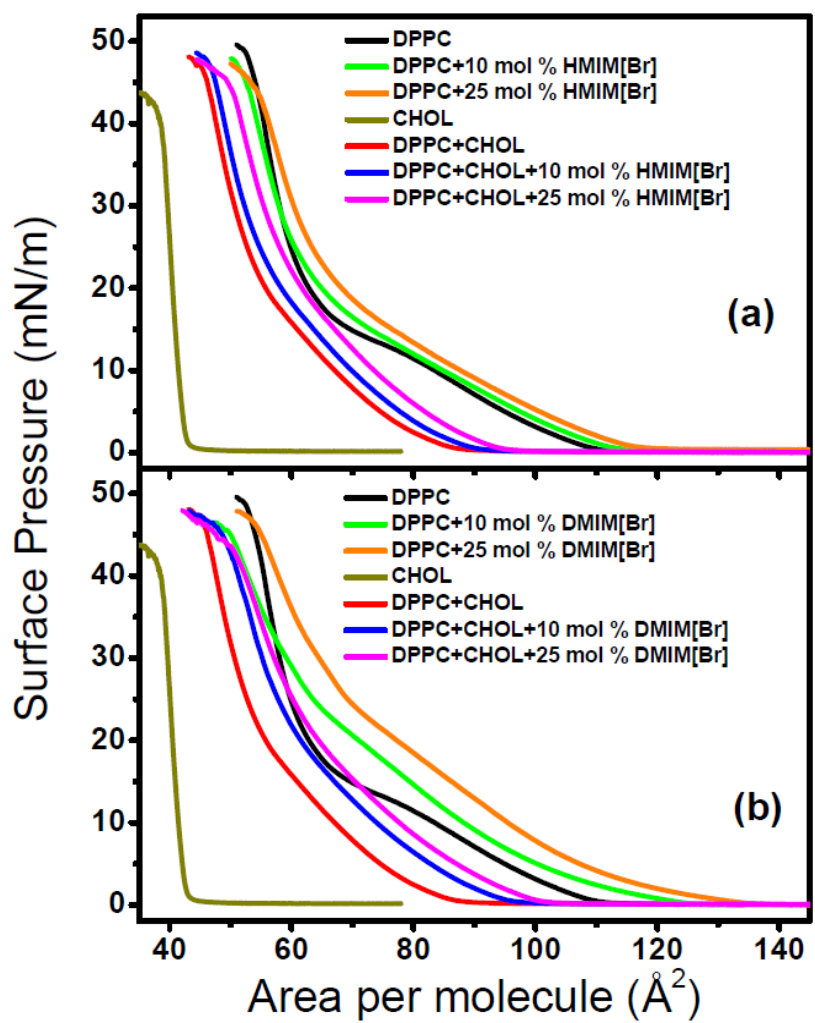

**FIGURE 2:** Surface pressure-area isotherms of monolayers formed at air-water interface, composed of pristine DPPC and DPPC+20 mol % cholesterol (CHOL) with added 10 and 25 mol % of ILs: (a) HMIM[Br] and (b) DMIM[Br]. All the measurements are performed at 300 K.

The alterations in APM are compared at two surface-pressures, namely, 5 mN/m (LE-phase) and 30 mN/m (LC-phase) and presented as bar chart in Fig. 3(a) and 3(b), respectively. As mentioned above, the addition of ILs in the pristine DPPC membrane led to the increase in APM and this effect was enhanced with the increase in ILs from 10 mol% to 25 mol%. A similar trend is noticed for DPPC/cholesterol mixed membrane in the presence of both ILs. Interestingly, the percentage of increase in APM is greater for DPPC/cholesterol mixed membrane as compared to pristine DPPC membrane. In LE phase (Fig. 3 (a)), the addition of short chain IL, HMIM[Br] in DPPC membrane increases the APM by 2.1% (10 mol%) and 5.2% (25 mol%) while, in DPPC/cholesterol membrane the APM increase by 4.2% (10 mol%) and 9.7% (25 mol%). Similarly, the addition of long chain IL, DMIM[Br] in DPPC membrane increases the APM by 5.4% (10 mol%) and 12.9% (25 mol%) while, in DPPC/cholesterol membrane, the APM increases by 11% (10 mol%) and 16.7% (25 mol%). Qualitatively similar responses are recorded in LC phase, as shown in Fig. 3(b). Such an observation of shifting the isotherm towards higher area with a higher interaction with increased alkyl chain length of ILs is reported earlier[23, 38]. The area occupied by a molecule is a critical factor that influences the thickness, flexibility and permeability of a lipid membrane. The impact of IL on this lateral area of molecule suggests that it could alter the physical properties of a cellular membrane, potentially affecting its functionality[39].

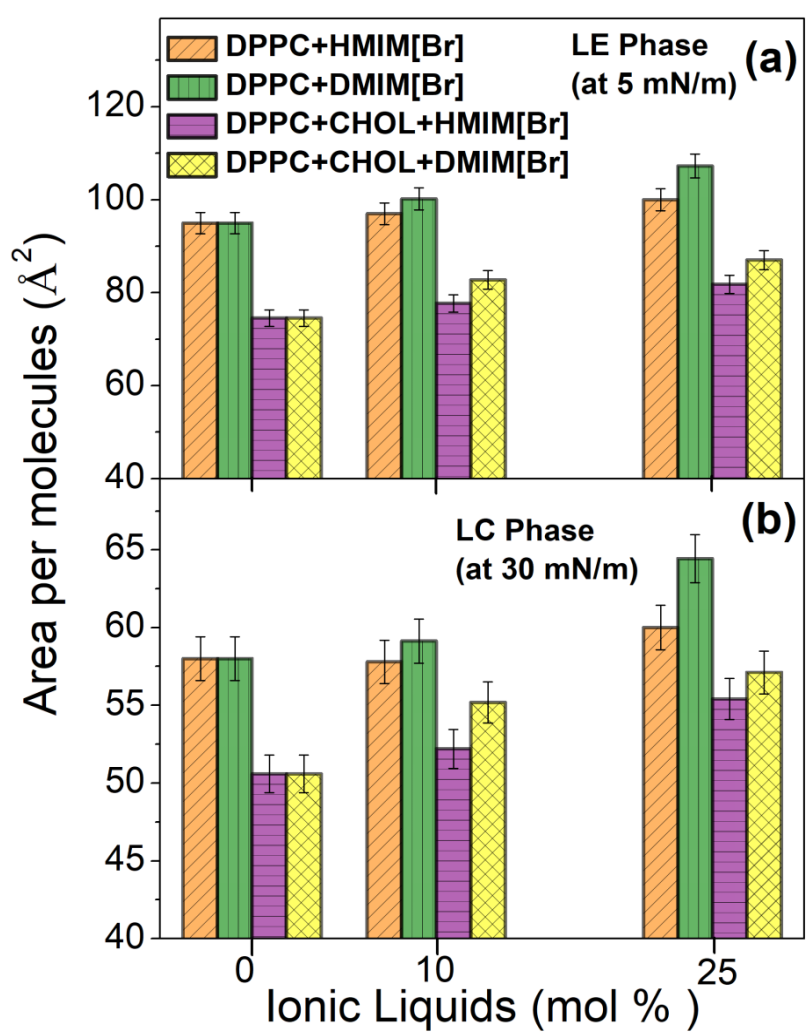

**FIGURE 3:** Bar chart to compare the area per molecule occupied by pristine DPPC and mixed DPPC/cholesterol lipid membranes with added 10 and 25 mol% of ILs, HMIM[Br] and DMIM[Br] at two surface pressures: (a) 5 mN/m denoting the LE-phase and (b) 30 mN/m denoting the LC-phase.

In the previous section, it has been shown that ILs show a prominent effect on the isotherm of the DPPC and mixed DPPC/cholesterol lipid monolayers. This interaction affects the properties of the membrane. One intriguing aspect is how ILs affect the viscoelasticity of the monolayer. Therefore, dilation rheology has been employed on the monolayer in the presence of ILs to analyze the dynamic viscoelastic nature of the lipid film. The measurements are performed to quantify the storage modulus ($E'$) and loss modulus ($E''$) in LE and LC phases, keeping the layer at a surface pressure of 5 and 30 mN/m, respectively. As shown in Fig. 4, the values of $E'$ is always greater than the $E''$ suggesting the elastic nature of lipid film. While $E'$ remains almost constant for both the phases, the value of $E''$ increases with the oscillatory frequency. The addition of cholesterol makes the lipids more ordered and, in turn, enhances the elasticity of the membrane. This is a well-known condensing effect of cholesterol[40]. Interestingly, the addition of short chain IL, HMIM[Br] in pristine DPPC layer increases its elasticity ($E'$) and viscosity ($E''$) significantly in both LE and LC phases as shown in Fig. 4 (a, b). The addition of this IL in mixed DPPC/cholesterol membrane reduces both moduli.

As shown in Fig. 4(c, d), the presence of DMIM[Br] in DPPC membrane reduces both the elasticity ($E'$) and viscosity ($E''$), making the lipid film more flexible. Addition of DMIM[Br] in DPPC/cholesterol membrane in LE phase also shows a reduction in both the $E'$ and $E''$ parameters. In the LC phase, $E'$ of mixed membrane increases and $E''$ decreases, qualitatively. A similar result for long chain IL, DMIM[Cl] has been reported for egg-sphingomyelin lipid membrane earlier[12]. It is further interesting to observe that the value of $E''$ also falls in the presence of the IL explaining a lower frictional force felt by a molecule. Such an observation accommodates well the higher area per lipid molecule observed in the surface-pressure area isotherm discussed in previous section. The lower value of $E''$ implies that the motion of the lipid molecules would be easier in the plane of a lipid layer.

Our measurements reflect that IL-membrane interaction strongly depend on the type of ILs and membrane composition. This is primarily dominated by electrostatic and hydrophobic interactions of the IL with the membrane assembly[34, 41]. Therefore, in the present study, HMIM[Br] and DMIM[Br] which differ in the alkyl chain length have shown different behavior while interacting with the lipid layer of varying composition. Shorter chain ILs mainly interact with the DPPC layer electrostatically placing themselves near the headgroup area of lipid. Effectively, the layer becomes positively charged which makes it harder to compress the film. It may lead to higher elasticity as quantified in the dilation rheology measurements. On the other hand, when this IL interacts with condensed DPPC/cholesterol assembly, it introduces higher APM and reverses the compacting effect of cholesterol leading to reduction in elasticity. In case of long chain DMIM[Br], due to higher hydrophobicity, it may enter the hydrophobic core of the lipid layer. Here, the effect of hydrophobicity may overshadow the electrostatic effect. Due to steric repulsion among the hydrocarbon chains, the APM increases more than the shorter chain one. It leads enhanced in-plane flexibility with lower elasticity. The reduction in elasticity due to DMIM[Br] is also observed in the DPPC/cholesterol film.

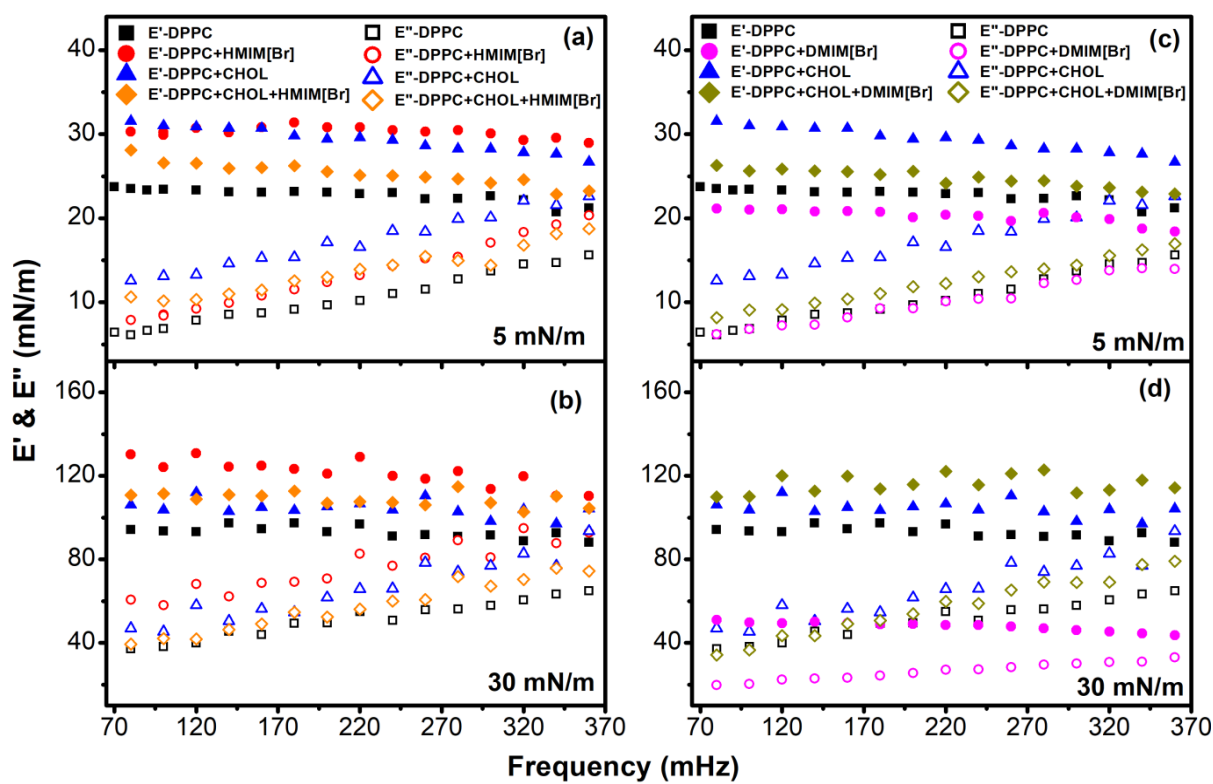

**FIGURE 4:** The storage ($E'$) and loss ($E''$) moduli of DPPC and DPPC/cholesterol mixed monolayer in the presence of 0 and 25 mol% of ILs at two different phases: (a) HMIM[Br] in liquid-expanded (LE) phase at 5 mN/m, (b) HMIM[Br] in liquid-condensed (LC) phase at 30 mN/m, (c) DMIM[Br] in liquid-expanded (LE) phase at 5 mN/m, and (d) DMIM[Br] in liquid-condensed (LC) phase at 30 mN/m.

The combined isotherm and rheology results provide compelling evidence of strong interactions between ILs and lipid monolayers, profoundly altering intermolecular interactions within the monolayer. Furthermore, cholesterol modulates IL-membrane interactions, with the effects varying based on the alkyl chain length of the IL. These findings highlight ILs as powerful modifiers of viscoelastic properties of lipid monolayers, emphasizing their pivotal role in shaping the mechanical behavior and structural integrity of membranes. Given that cell membrane consists of two opposing monolayers forming a bilayer structure, we further investigate the impact of ILs on the DPPC and DPPC/cholesterol bilayers.

**4.2 Role of cholesterol on the effect of ILs on the phase behavior of lipid bilayer**

The impact of cholesterol on IL-membrane interactions was investigated by examining how it modulates the effects of ILs on the phase behavior of DPPC membranes, using DSC and FTIR techniques. These two methods complement each other, offering microscopic insights into the membrane's phase behavior. DSC enables us to measure changes in the enthalpy of the membrane during phase transitions[42]. FTIR captures IR-active bands and fundamental vibrations of the $CH_2$ group (e.g., C-H stretching) which are highly sensitive to the *trans-gauche* ratio[43]. Tracking these bands is ideal for studying the main phase transition of the membranes, which particularly involves alkyl tails transforming from an all-*trans* state to a disordered state with significant *gauche* defects (conformational disorder). Therefore, FTIR measurements furnish valuable information about the conformation of lipid molecules during the phase transition, serving as a valuable complement to the insights gained from DSC results. Observed DSC thermograms for cholesterol-enriched DPPC membrane and pure DPPC membrane in the absence and presence of HMIM[Br] and DMIM[Br] are shown in the Fig. 5 (a) & (b), respectively. While analyzing the DSC thermograms, three parameters namely (i) transition temperature (onset or peak position), (ii) enthalpy associated with the transition, and (iii) peak width are important for characterizing IL-lipid membrane interactions. Incorporation of cholesterol significantly alters the thermotropic phase behavior of DPPC membranes. In the DPPC–cholesterol system, the pre-transition disappears entirely, and the main phase transition becomes markedly broader (Fig. S1). Moreover, the enthalpy associated with the main transition is substantially reduced to 23 kJ/mol, compared to pure DPPC (32 kJ/mol). Cholesterol is well known to induce a liquid-ordered ($L_o$) phase, which exhibits intermediate characteristics between the highly ordered gel phase and the highly mobile liquid-disordered ($L_d$) phase. In this state, phospholipid chains maintain a high degree of order akin to the gel phase while simultaneously displaying rapid lateral mobility, a hallmark of the $L_d$ phase. Upon cholesterol addition, both the onset temperature ($T_{on}$) and the peak transition temperature ($T_{peak}$) shift to lower values, with a more pronounced decrease in $T_{on}$, leading to an overall broadening of the phase transition peak. The introduction of ILs into DPPC–cholesterol membranes further modifies this behavior as evident from Fig.5(a). Both HMIM[Br] and DMIM[Br] shift the main phase transition to lower temperatures, with the longer-chain IL, DMIM[Br], having a more pronounced effect. Similarly, $T_{on}$ also decreases upon IL incorporation, though the change is less significant for the shorter-chain IL. The broadening of the main transition peak is particularly notable with DMIM[Br], although no significant change in transition enthalpy is observed. In the case of pristine DPPC, the DSC thermogram displays a weak and broad pre-transition peak (FWHM = 4 K) around 308 K and a strong, sharp main transition peak (FWHM = 0.4 K) at 315 K, corresponding to enthalpies of 2 and 32 kJ/mol, respectively, as shown in Fig. 5(b)[24]. Upon incorporation of HMIM[Br], the pre-transition remains visible, while the main transition shifts slightly to a lower temperature and the peak broadens (FWHM increases from 0.4 to 0.6 K), though its enthalpy remains unchanged. In contrast, the longer-chain IL, DMIM[Br], induces more pronounced effects: the pre-transition disappears entirely, and the main transition shifts significantly to lower temperature with substantial peak broadening (FWHM = 1.6 K). Despite these changes, the transition enthalpy remains approximately 32 kJ/mol, indicating that single-chain ILs influence membrane fluidity and cooperativity without altering the total energy involved in the transition[24]. Interestingly, double-chain ILs such as 1-dodecyl-3-methyl-imidazolium bromide (DDMIM[Br]) affect both the transition temperature and enthalpy[43]. Unlike single-chain ILs, which intercalate between lipid acyl chains and reduce $T_{peak}$ without altering ΔH, double-chain ILs displace lipid molecules within the membrane. This displacement leads to a reduction in both the number of lipids participating in the transition and the associated enthalpy. Therefore, these contrasting effects underscore fundamental differences in membrane–IL interaction mechanisms between single-chain and double-chain ILs.

    For quantitative comparison, the obtained $T_{peak}$ and FWHM for DPPC and DPPC-cholesterol membranes, in the absence and presence of ILs are shown in Fig.5 (c) & 5 (d), respectively. The incorporation of ILs shifts the main phase transition to lower temperatures in both membrane systems, with and without cholesterol. Notably, ILs with longer alkyl chains induce more pronounced effects as shown in Fig. 5 (c). Incorporating DMIM[Br] induces a 9 % shift in $T_{peak}$ of pure DPPC, which increases to 11% in the DPPC-cholesterol membrane. As evident from Fig. 5 (d), the addition of DMIM[Br] leads to a substantial increase of approximately 300% in the FWHM of the peak corresponding to the main phase transition for pure DPPC, whereas for DPPC with cholesterol, where the peak is already broadened, the relative change in FWHM is comparatively modest at around 20%. Our DSC measurements suggest that the effects of ILs on the phase behavior of DPPC and DPPC-cholesterol membranes are qualitatively similar.

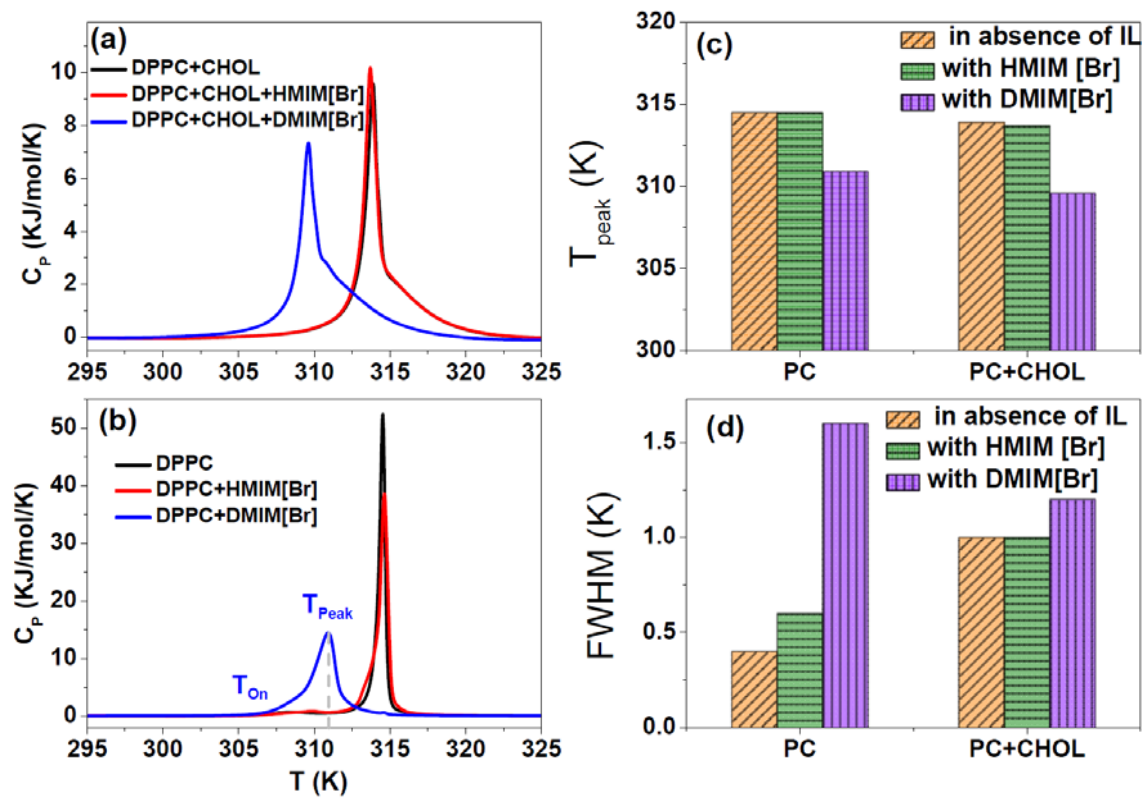

**FIGURE 5:** DSC thermograms of ULVs composed of: (a) DPPC with cholesterol and (b) DPPC, in the absence and presence of 50 mol % HMIM[Br] and DMIM[Br] during heating cycles. Onset ($T_{on}$) and peak ($T_{peak}$) transition temperatures are indicated for a representative DSC thermogram. Impact of incorporating HMIM[Br] and DMIM[Br] on (c) $T_{peak}$, and (d) FWHM corresponding to the main transition temperature for DPPC and DPPC with cholesterol

**Role of cholesterol on the effect of ILs on the conformation of the lipid**

To further investigate phase behaviour, we conducted FTIR measurements to study the conformational changes in the membrane. Cholesterol modulates the conformation of the lipid molecules via condensing effects, indicating that the DPPC-cholesterol membrane has an intrinsically different conformational order compared to pure DPPC membrane. Therefore, it is imperative to investigate the effects that ILs elicit on the conformational ordering of DPPC-cholesterol membranes as opposed to pure DPPC membranes. FTIR measurements on DPPC and DPPC-cholesterol membranes in the absence and presence of ILs not only provide information about the impact of ILs on the conformational of lipids but also provide microscopic insights into the phase transitions in the membranes. The structure of DPPC molecule comprises two hydrocarbon alkyl chains as shown in Fig.1. In the FTIR measurements, we focus on $CH_2$ symmetric stretching mode ($\upsilon_s(CH_2)$) present in the alkyl chains of DPPC lipid to monitor changes in the tail region. This $\upsilon_s(CH_2)$ mode serves as an indicator of *trans-gauche* isomerization in the alkyl chains, offering valuable insights into the conformational ordering/disordering in both the membranes[43].

The temperature-dependent FTIR absorbance spectra of the $\upsilon_s(CH_2)$ mode within the alkyl chains of DPPC membrane with and without 50 mol% of HMIM[Br] and DMIM[Br] ILs, are shown in Fig. S2 (a-c). The spectral analysis spans the wave number range of 2830-2870 $cm^{-1}$. At low temperature, the absorbance spectra of pure DPPC membrane are centered at ~ 2850 $cm^{-1}$. Beyond 315 K, a noticeable sharp blueshift (2 $cm^{-1}$) in absorbance spectra is evident, as shown in Fig. S2 (a). The incorporation of HMIM[Br] into the DPPC membrane enhances this blueshift to 2.7 $cm^{-1}$, as shown in Fig S2 (b). Furthermore, the temperature, at which this blueshift occurs, decreases to 314 K. Upon incorporating DMIM[Br] into the DPPC membrane, this blueshift further enhances to 2.9 $cm^{-1}$, as shown in Fig. S2 (c). Additionally, the temperature, at which this shift occurs, further decreases to 310 K.

The temperature-dependent FTIR absorbance spectra of the $\upsilon_s(CH_2)$ mode within the alkyl chains of DPPC in the presence of cholesterol, both with and without 50 mol% of HMIM[Br] and DMIM[Br], are shown in Fig. S2 (d-f). In the absorbance spectra of DPPC membrane with cholesterol, no sharp blueshift is observed with increasing temperature, as shown in Fig. S2 (d). However, the onset of a noticeable shift in spectra becomes evident at ~ 310 K. This is consistent with our DSC measurements which suggest that the presence of cholesterol broadens the main phase transition and the onset temperature shifts towards a lower value. The incorporation of HMIM[Br] into the DPPC with cholesterol membrane does not result in a significant change in the blueshift temperature onset, as evident in Fig. S2 (e). However, with the presence of DMIM[Br] in the DPPC with cholesterol membrane, the onset of the blueshift temperature mildly decreases to 309 K, as evident in Fig. S2 (f).

The variations in the peak centers of the $\nu_s(CH_2)$ mode are indicative of changes in the *gauche*/*trans* ratio within lipid chains, providing insights into the order/disorder state that affects the phase transition behavior of lipid membranes[43]. The phase behavior of pure DPPC membrane as studied using FTIR is discussed earlier[23]. The role of cholesterol is crucial in influencing the phase transition and phase behavior of DPPC membrane. The temperature-dependent peak centers of $\nu_s(CH_2)$ mode for DPPC membrane in the presence of cholesterol are shown in Fig S3. For comparison, temperature-dependent peak centers of $\nu_s(CH_2)$ mode for pure DPPC membrane is also shown in the same Figure. Remarkably, at 305 K, the observed peak centers for DPPC-cholesterol membrane are approximately 2850.7 cm$^{-1}$, closely resemble those in pure DPPC membrane. This peak position indicates a predominance of all-*trans* conformations in the alkyl chains, suggesting that the presence of cholesterol does not substantially alter chain ordering at this temperature. Therefore, at this temperature, the presence of cholesterol maintains the $L_{\beta'}$ phase in DPPC membrane, which aligns with findings of earlier studies[18, 44]. However, as the temperature increases, a subtle blueshift in peak centers is observed in the DPPC-cholesterol membrane, in contrast to the sharp blueshift at 315 K ($T_m$) observed in the pure DPPC membrane. This subtle blueshift indicates that the presence of cholesterol induces mild *gauche* defects in the alkyl chains of the DPPC membrane. The rise in *gauche* defects suggests that the presence of cholesterol introduces disorder in the gel phase of DPPC membrane and adopts the liquid-ordered phase ($L_{\beta'}+L_o$). The presence of cholesterol leads to a substantial decrease in the onset and a significant broadening of the main transition. In the $L_\alpha$ phase (above $T_m$), presence of cholesterol, compared to pure DPPC membrane, induces a redshift in peak centers, implying a decrease in the *gauche/trans* ratio i.e., increase in the alkyl chain order of lipids and membrane adopts the $L_\alpha + L_o$ phase. The introduction of cholesterol to lipid bilayers leads to molecular re-organization causing a condensing effect which leads to a reduction in the surface area or increase in the alkyl chain order of lipid[18]. This condensing effect of cholesterol is universal for all the types of membrane[45].

The temperature-dependent peak centers of the $\nu_s(CH_2)$ mode for DPPC membranes with cholesterol and for pure DPPC membrane, in absence and presence of ILs, are shown in Fig. 6(a) and (b), respectively. In Fig. 6 (a), the presence of HMIM[Br] exhibits a modest effect on the conformation of lipids and the onset of main transition temperature of the DPPC with cholesterol membrane. In the $L_\alpha + L_o$ phase, a blueshift in the peak centers is observed due to the presence of HMIM[Br]. In case of DMIM[Br], a more pronounced blueshift is observed in both the phases of DPPC with cholesterol membrane. In case of pure DPPC, addition of HMIM[Br] causes a slight blueshift in $\nu_s(CH_2)$ peaks for both the phases and reduces $T_m$ to 314 K as evident from Fig. 6 (b). The introduction of DMIM[Br] results in a more significant decrease in $T_m$ to 309 K. Additionally, larger blueshifts in peak centers are observed in both phases with DMIM[Br] compared to HMIM[Br] at the same concentration.

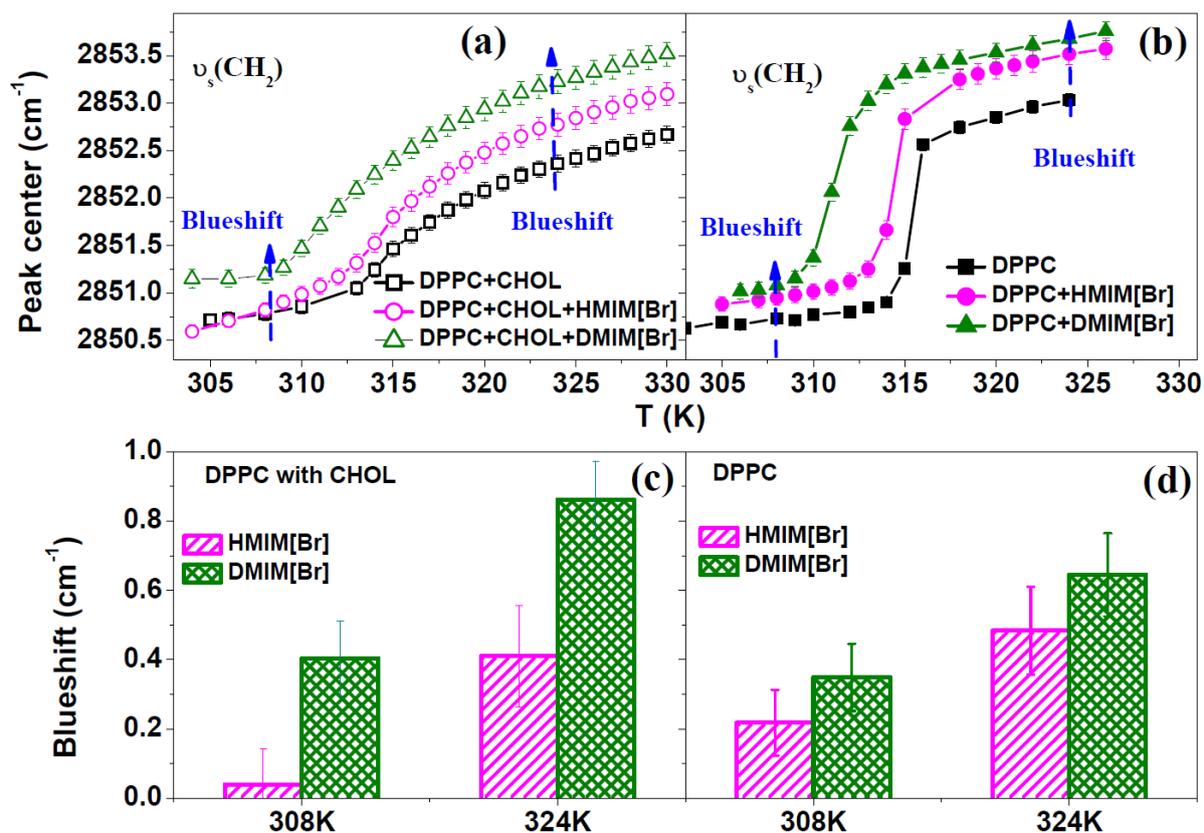

**FIGURE 6:** Temperature-dependent peak centers of $\nu_s(CH_2)$ mode in the alkyl chains of (a) DPPC supplemented with cholesterol and (b) DPPC, with and without 50 mol % HMIM[Br] and DMIM[Br]. The blueshift in peak centers of $\nu_s(CH_2)$ mode in (c) DPPC enriched with cholesterol and (d) DPPC due to the presence of HMIM[Br] and DMIM[Br] at 308 K (gel phase) and 324 K (fluid phase).

In Fig. 6 (c) and (d), the blueshift, as indicated by the blue arrows in Fig. 6 (a) and (b), resulting from the addition of ILs is illustrated for both DPPC-cholesterol and pure DPPC membranes. This depiction serves as a quantitative comparison of the observed blueshift in $\nu_s(CH_2)$, highlighting the impact of ILs on both types of membranes. It is evident that the incorporation of ILs leads to an increase in *gauche* conformations for both membrane systems. However, the magnitude of the blueshift depends on the physical state, composition of the membrane, as well as the chain length of ILs. The presence of cholesterol in the membrane is observed to modulate the magnitude of blueshift induced by ILs. Specifically, with HMIM[Br], the blueshift in the case of DPPC-cholesterol membrane is relatively smaller than in the pure DPPC membrane. Conversely, the incorporation of DMIM[Br] into the DPPC-cholesterol membrane shows a substantial blueshift, which is in fact higher than that observed in the pure DPPC membrane. This suggests that the perturbing effect of DMIM[Br] on the DPPC-cholesterol membrane is more pronounced in comparison to the pure DPPC membrane. These measurements indicate that the role of cholesterol in the IL-membrane interaction is not uniform but rather depends on the specific IL-membrane systems chosen. The particular difference observed in this study may be attributed to the longer alkyl chain length of DMIM[Br], which exerts a more pronounced perturbing effect in the case of DPPC-cholesterol membrane. These measurements support the perturbation and softening in the membrane observed by DSC and other techniques.

The FTIR absorbance peak bandwidth of the $\nu_s(CH_2)$ mode offers insights into the rotational dynamics of alkyl chains within the DPPC membrane[46]. Figure S4 (a) & (b) present temperature-dependent variations in the half width at half maxima (HWHM) of the $\nu_s(CH_2)$ mode, in the presence of ILs in pure DPPC and DPPC-cholesterol membranes, respectively. In the pure DPPC membrane, as temperature rises, the HWHM increases, reflecting enhanced rotational motion and increased fluidity. A significant jump at 315 K suggests accelerated chain mobility due to main phase transition. Incorporating HMIM[Br] consistently increases the HWHM, indicating enhanced alkyl chain rotation and augmented fluidity in both phases. Augmenting of the rotational dynamics is substantially enhanced for the case of DMIM[Br] suggesting that incorporation of longer chain length IL significantly boosts the lipid rotation. This underscores the role of IL alkyl chain length in influencing rotational dynamics and overall membrane fluidity, aligning with previous studies[5, 47], highlighting that IL's impact on lipid motion is accentuated with longer IL chain lengths. In the case of DPPC-cholesterol membrane, at low temperatures below the $T_m$, a larger HWHM compared to pure DPPC membrane implies increased alkyl chain rotation due to presence of cholesterol. As temperature increases, the HWHM continues to increase. Above the main phase transition, HWHM for DPPC membrane with cholesterol remains lower than that in the pure DPPC membrane, indicating reduced membrane dynamics with cholesterol. Interestingly, HMIM[Br] shows negligible impact on HWHM, while DMIM[Br] enhances HWHM, indicating increased dynamics in the membrane supplemented with cholesterol. FTIR measurements further suggest that the perturbing effects of HMIM[Br] are diminished due to the presence of cholesterol, while the effects of DMIM[Br] are enhanced.

**Role of cholesterol on the binding affinity of IL with lipid membrane**

To probe the impact of cholesterol on the binding affinity of DMIM[Br] with the lipid membrane, we have carried out ITC measurements. These experiments not only generate binding isotherms but also provide a thorough thermodynamic understanding of the interaction between the ILs and the membrane. The titration of DMIM[Br] into the DPPC membrane with cholesterol yields isotherms, indicating an exothermic interaction at 320 K, as illustrated in Fig. 7. Successful injections showcase a reduction in enthalpy as the IL concentration increased. The binding isotherm of DMIM[Br] with the pure DPPC membrane is included for direct comparison[5]. Notably, the binding isotherm for the pure DPPC membrane exhibited a rapid rise and saturation at a significantly lower IL/lipid ratio than the DPPC-cholesterol membrane. This reflects that the membrane composition influences the binding affinity of the DMIM[Br] and is notably weaker in the presence of cholesterol. While we do not anticipate substantial differences in the electrostatic or hydrophobic interaction of IL with the lipid membrane in both scenarios, the primary distinction likely arises from variations in the physical properties of the membranes. Cholesterol is known for its condensing effect on the membrane, which imparts rigidity and compactness. In this scenario, lipid molecules are organized and densely packed, impeding the insertion of IL into the membrane.

Single site binding model was used to describe the observed thermograms and association constants ($K_a$) governing the interaction between DMIM[Br] and DPPC membranes with cholesterol are determined to be $1.9 \times 10^3$ M$^{-1}$, marking an approximate fivefold reduction compared to the association constant for the interaction between DMIM[Br] and pure DPPC membranes ($9.4 \times 10^3$ M$^{-1}$)[5]. This association constant facilitates the computation of the Gibbs free energy associated with the binding of DMIM[Br] to the lipid membrane, utilizing the formula $\Delta G = -RT \ln 55.5 K$, where 55.5 M denotes the concentration of water. The calculated ΔG for DMIM[Br] is observed to be -8.4 kcal/mol and -7.3 kcal/mol with pure DPPC and DPPC membranes with cholesterol, respectively. Clearly, in the presence of cholesterol, both the association constant and the associated Gibbs free energy exhibit a decrease when compared to the pure DPPC membrane. This suggests that the presence of cholesterol weakens the binding of IL with the lipid membrane. This is likely because of the stiffened cholesterol + DPPC membranes tend to avoid incursion of the IL molecules.

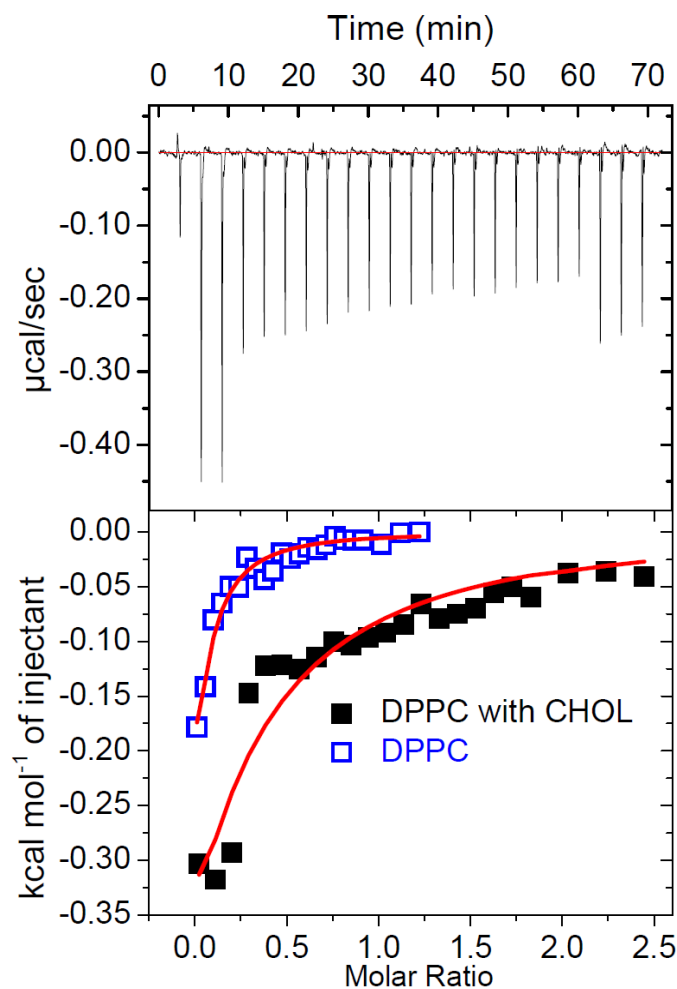

**FIGURE 7:** Isothermal titration calorimeter (ITC) traces and thermograms depicting the interaction of DMIM[Br] with DPPC containing cholesterol at 320 K. To facilitate a direct comparison, thermogram of DMIM[Br] interacting with pure DPPC[5] at 320 K is also presented.

**Role of cholesterol on the impact of IL on the thickness of lipid membrane**

Our recent SANS study revealed that interaction of ILs with DPPC membrane depends on the alkyl chain length of ILs[24]. The shorter-chain IL, HMIM[Br], induces the aggregation of DPPC ULVs. In contrast, this aggregation is absent with the longer alkyl chain IL, DMIM[Br]. However, both ILs leads to bilayer thinning of DPPC membrane[24]. SANS measurements were carried out to investigate how the presence of cholesterol affects ILs effects on the ULVs. Figure 8 displays SANS profiles for ULVs composed of DPPC with cholesterol, at varying concentrations of HMIM[Br] and DMIM[Br]. SANS data on pure DPPC ULVs at different concentrations of both ILs are also shown for direct comparison[24].

In the absence of ILs, scattering intensity for both DPPC and DPPC with cholesterol showed $Q^{-2}$ dependence in the low-$Q$ region, and the absence of Bragg peaks confirmed successful ULVs formation. The lack of a low-$Q$ cut-off indicated that the ULVs size exceeded the detectable $Q$ range of our measurements.

Similar to pure DPPC[24], the form factor is utilized for the analysis of DPPC supplemented with cholesterol ULVs. Using a core-shell model for analysis, we fitted the SANS data with a fixed ULV radius ($R$), a value greater than $2\pi/Q_{min}$ (~ 100 nm), providing insights into bilayer thickness. The extracted parameters from the fitting are summarized in Table-1. For direct comparison, parameters obtained for pure DPPC at different concentration of ILs are given in Table-1[24]. At 300 K, the bilayer thickness of pure DPPC ULVs is measured to be 41.5 Å. With increasing temperature, at 330 K, a reduction of ~ 6 Å in bilayer thickness is observed. FTIR results suggest that at higher temperature, the alkyl chains of the lipids become disordered, displaying significant *gauche* defects, which contribute to the thinning of the bilayer. Cholesterol exerts a minimal effect on bilayer thickness in the gel phase but induces notable thickening in the fluid phase. This behavior arises from the combined influence of acyl chain ordering and tilt angle adjustments[48]. In gel phases, incorporation of cholesterol leads to a reduction in the acyl chain tilt angle which contributes to an increase in bilayer thickness. However, cholesterol's impact on acyl chain ordering differs markedly between gel and fluid phases. In the gel phase, it introduces acyl chain disorder, while in the fluid phase, it enhances chain ordering[49]. Consequently, in the gel phase, the opposing effects of reduced tilt and increased disorder counteract each other, resulting in negligible changes to bilayer thickness. In contrast, in the fluid phase, enhanced ordering led to a pronounced increase in bilayer thickness. These findings are corroborated by FTIR results (Fig S3), which show no significant shift in the $v_s(CH_2)$ peak position in the gel phase, indicating minimal change in chain ordering, but reveal enhanced ordering in the fluid phase in the presence of cholesterol.

Recent study has suggested that shorter-chain IL HMIM[Br] induces aggregation in DPPC vesicles[24]. However, as shown in Fig. 8(a), the presence of cholesterol significantly mitigates the bilayer aggregation induced by the HMIM[Br]. Aggregation is observed only at higher concentrations of HMIM[Br], and exclusively at the lower temperature of 300 K, where the DPPC membrane is in the gel phase. Even then, the aggregates appear to be more loosely bound, as indicated by a reduced volume fraction and increased center-to-center distance between bilayers compared to the pure DPPC system. At 330 K, no aggregation is observed, further highlighting cholesterol's stabilizing effect. In the case of DMIM[Br], no aggregation is detected in the cholesterol-containing DPPC ULVs across all concentrations and temperatures (Fig. 8b), suggesting that the longer alkyl chain of DMIM[Br] does not induce significant membrane perturbation under the conditions studied. In contrast, in the absence of cholesterol, the incorporation of HMIM[Br] leads to pronounced bilayer aggregation in DPPC ULVs, as seen in Fig. 8(c) [24]. The scattering profiles show a clear shift in Q-dependence from $Q^{-2}$ to nearly $Q^{-3}$, which is indicative of aggregate formation[24]. To capture this behavior, the SANS data were fitted using a model that includes both the ULV form factor P(Q) and a structure factor S(Q) based on Baxter's sticky hard sphere model to account for inter-vesicle interactions. The fitting parameters are listed in Table 1. A comparison between 300 K and 330 K reveals that at the higher temperature, the volume fraction decreases and the center-to-center distance between bilayers increases for both HMIM[Br] concentrations. This likely reflects the effect of thermal energy reducing IL-induced aggregation by loosening the vesicle packing. Notably, in the case of DMIM[Br], no aggregation is observed even in pure DPPC ULVs (Fig. 8d), and the scattering profiles remain largely unchanged, indicating that the longer alkyl chain limits inter-vesicle interactions and aggregation.

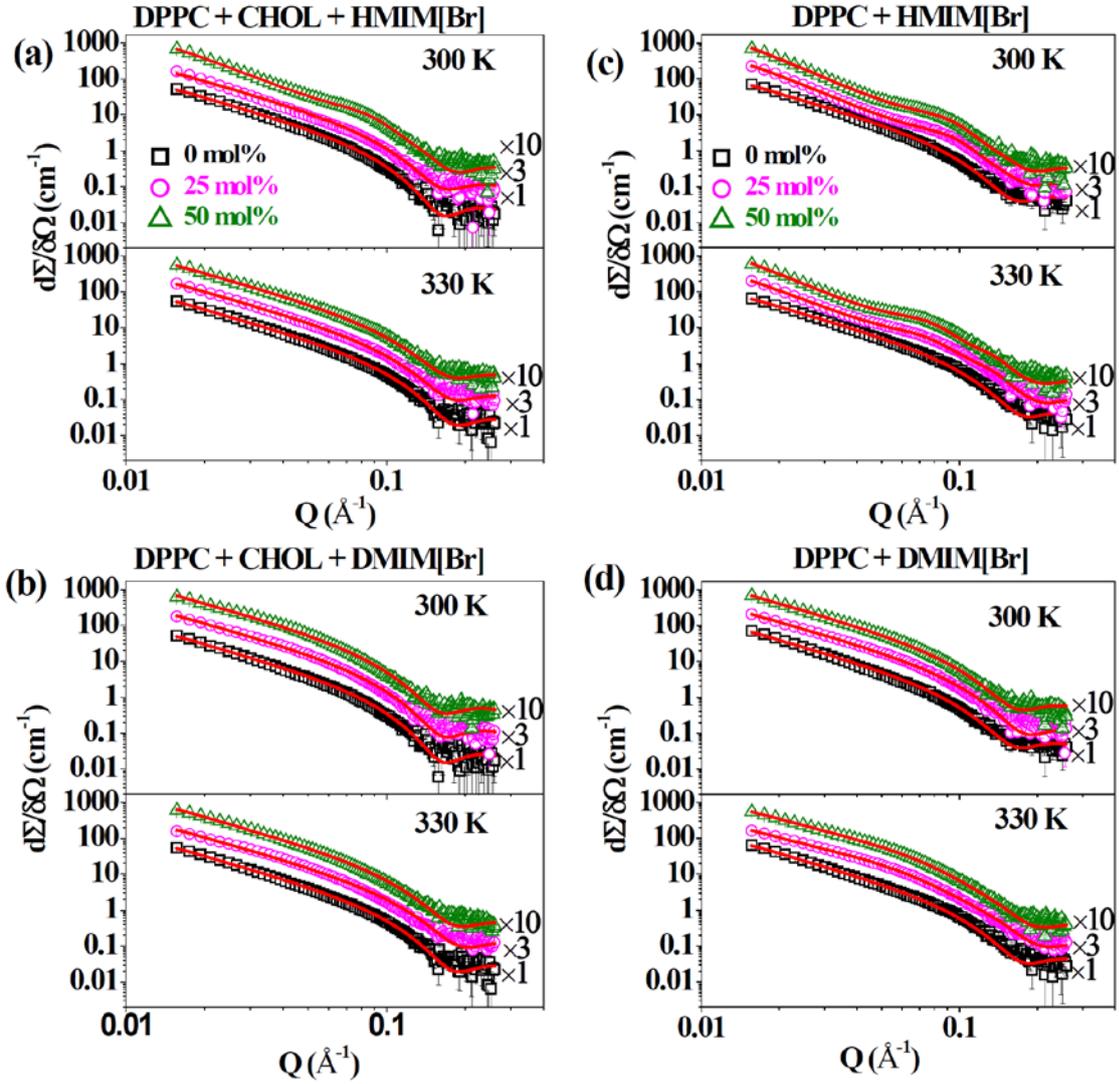

**FIGURE 8:** SANS data profiles for ULVs composed of DPPC with cholesterol at 0 mol% (black), 25 mol% (magenta), and 50 mol% (green) of (a) HMIM[Br] and (b) DMIM[Br] at temperatures of 300 K and 330 K. For direct comparison, SANS data for cholesterol free ULVs at 300 K and 330 K at different concentrations of (c) HMIM[Br] and (d) DMIM[Br] are adopted from Ref[24] Fits are shown by the solid red lines.

**TABLE 1: Variations in bilayer thickness, volume fractions, and center-to-center distance of DPPC with cholesterol bilayers with differing concentrations of HMIM[Br] and DMIM[Br] at 300 K and 330 K. For direct comparison, these parameters for DPPC bilayer with different concentrations of ILs are also given[24].**

| Name of IL | IL conc. (mol%) | T (K) | DPPC + cholesterol | | | DPPC | | |
|---|---|---|---|---|---|---|---|---|
| | | | Bilayer thickness (Å) | Volume fraction (×10⁻²) | Center-to-center distance of bilayer | Bilayer thickness (Å) | Volume fraction (×10⁻²) | Center-to-center distance of bilayer |
| - | 0 | 300 | 41.4± 1.3 | - | - | 41.5 ± 1.5 | - | - |
| | | 330 | 36.9± 1.0 | - | - | 35.4 ± 1.0 | - | - |
| HMIM[Br] | 25 | 300 | 38.0± 1.0 | - | - | 35.3 ± 1.0 | 6.1 ± 0.4 | 40.8 ± 1.0 |
| | | 330 | 34.6± 0.7 | - | - | 30.8 ± 0.6 | 4.0 ± 0.3 | 50.3 ± 1.8 |
| | 50 | 300 | 34.1± 0.8 | 4.1 ± 0.3 | 43.2± 1.6 | 34.3 ± 0.8 | 5.3 ± 0.3 | 41.5 ± 1.5 |
| | | 330 | 33.1± 0.8 | - | - | 32.0 ± 0.8 | 5.0 ± 0.3 | 49.3 ± 1.5 |
| DMIM[Br] | 25 | 300 | 38.5± 1.0 | - | - | 35.1 ± 1.0 | - | - |
| | | 330 | 33.1± 0.8 | - | - | 30.7 ± 0.6 | - | - |
| | 50 | 300 | 36.0± 0.8 | - | - | 34.1 ± 0.8 | - | - |
| | | 330 | 33.5± 0.8 | - | - | 31.2 ± 0.7 | - | - |

Significant bilayer thinning was observed in DPPC membranes, both with and without cholesterol, upon the addition of ILs, as detailed in Table 1. The inclusion of HMIM[Br] reduced the bilayer thickness to approximately 35.3 Å in the gel phase and 30.8 Å in the fluid phase. This thinning effect is attributed to the combined influence of the IL in inducing gauche defects in the lipid alkyl chains, as confirmed by FTIR measurements, and the interdigitation of lipids, as observed through X-ray reflectivity[24, 43], as well as the interdigitation of lipids. As the concentration of HMIM[Br] increases, these *gauche* defects and interdigitation between acyl chains become more pronounced, leading to further bilayer thinning in both phases. In comparison, DMIM[Br] induces a more substantial reduction in bilayer thickness than HMIM[Br]. This enhanced effect arises because the longer alkyl chain of DMIM[Br] penetrates deeper into the lipid bilayer, causing greater disruption to the hydrophobic core and resulting in more pronounced thinning compared to shorter-chain ILs. The presence of cholesterol mitigates bilayer thinning due to the incorporation of ILs. This is consistent with ITC results, which demonstrate a relatively weaker binding affinity of ILs to DPPC membranes containing cholesterol compared to those with pure DPPC. These observations highlight the significant role of cholesterol and alkyl chain length of IL in modulating IL-membrane interactions and structural organization.

**Role of cholesterol on the effects of IL on the size of vesicles**

While SANS hints at ULVs aggregation at higher concentration of HMIM[Br], the limited *Q*-window of the experiment restricts the determination of overall size. Therefore, DLS measurements were performed to assess ULV sizes and confirm their aggregation. Figure S5 presents the autocorrelation functions (ACFs) for DPPC ULVs with cholesterol across varying concentrations of HMIM[Br] and DMIM[Br] at 300 and 330 K. For direct comparison, the ACFs of pure DPPC ULVs at various IL concentrations are also presented in Fig. S5[24]. The data highlights a stark difference in the effects of ILs on ACFs in the presence and absence of cholesterol. When cholesterol is incorporated into DPPC ULVs, the impact of ILs on vesicle dynamics is significantly attenuated. For HMIM[Br], only a slight slowing of the ACF decay is observed at 300 K, even at higher concentrations (Fig. S5a), suggesting a minimal increase in vesicle size and a relatively muted effect on diffusion. At 330 K, the influence of HMIM[Br] is negligible. In the case of DMIM[Br], the presence of cholesterol further diminishes any observable effects; across all concentrations and temperatures, the ACF decay remains largely unchanged (Fig. S5b), indicating that cholesterol stabilizes the vesicle structure and suppresses IL-induced perturbations. This is consistent with the SANS results. In contrast, for pure DPPC ULVs (without cholesterol), the addition of HMIM[Br] at 300 K results in a markedly slower ACF decay (Fig. S5c), with the effect becoming more pronounced at higher IL concentrations (up to 50 mol%). This slower decay suggests a reduction in the diffusion coefficient, likely due to an increase in ULV size. Moreover, at elevated HMIM[Br] concentrations, the ACF decay shifts from unimodal to bimodal, indicating more complex diffusion behavior, possibly due to vesicle aggregation or heterogeneous size distributions. At 330 K, the influence of HMIM[Br] is less significant, with noticeable delays in ACF decay observed only at higher concentrations (Fig. S5c). In comparison, DMIM[Br] exhibits a different pattern. At 300 K, it induces no substantial change in ACF decay across all concentrations (Fig. S5d), while at 330 K, a slight delay emerges, suggesting minor effects on ULV diffusion at elevated temperatures.

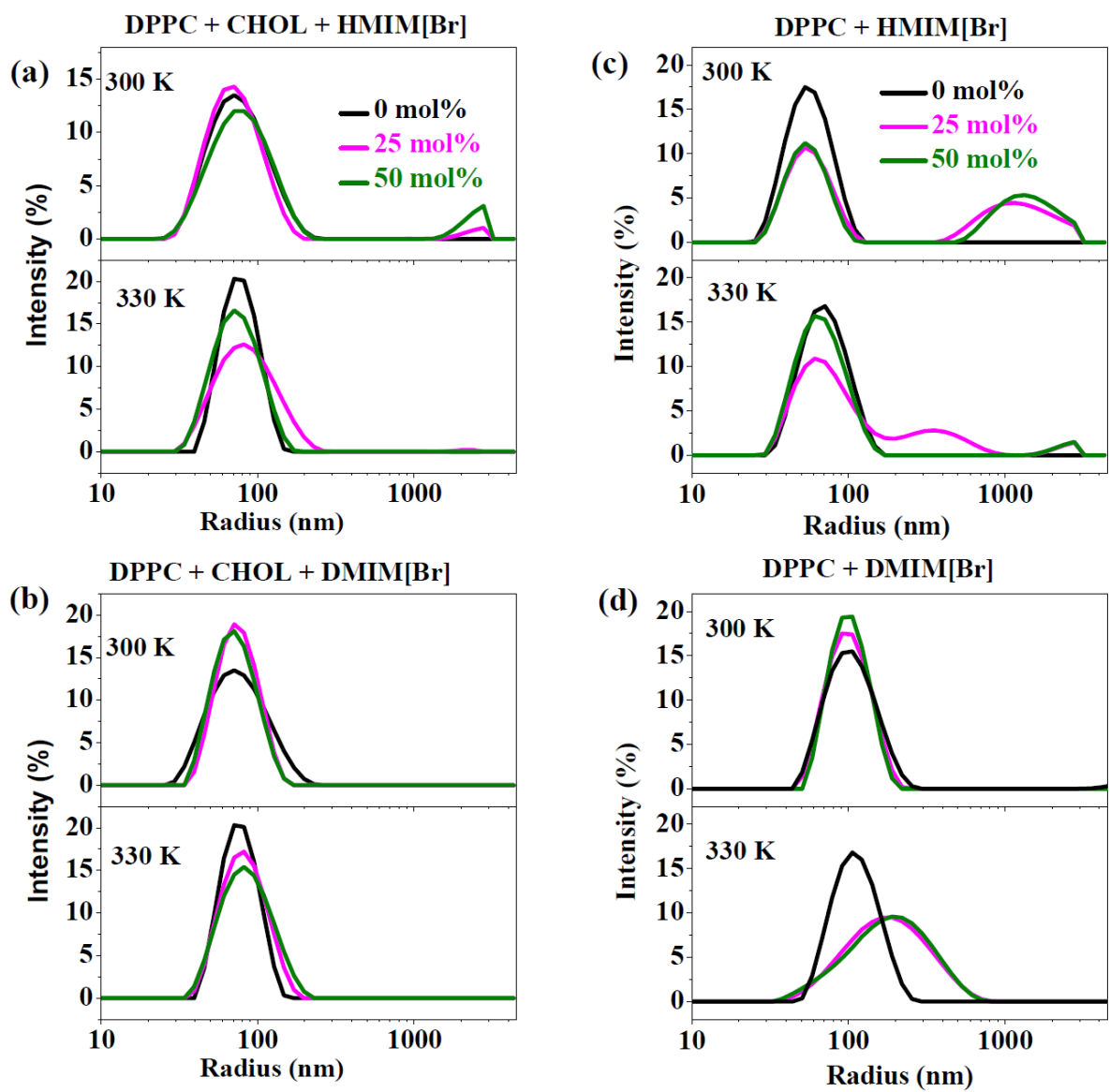

**FIGURE 9:** The size distributions of DPPC with cholesterol ULVs at different concentrations of (a) HMIM[Br] and (b) DMIM[Br]. For direct comparison, the distribution profile for DPPC ULVs at different concentrations of (c) HMIM[Br] and (d) DMIM[Br] are taken from Ref[24].

Due to the higher polydispersity index (PDI > 0.1) of vesicles in the presence of ILs, the CONTIN method[50] was employed for data analysis. The intensity-weighted size distributions of DPPC+Choletersol and DPPC[24] ULVs with varying IL concentrations and chain lengths are shown in Fig. 9. It is evident that the inclusion of cholesterol markedly alters the interaction of ILs with DPPC ULVs, effectively mitigating IL-induced aggregation. As shown in Fig. 9(a-b), cholesterol-containing DPPC ULVs exhibit negligible changes in size distribution upon addition of either HMIM[Br] or DMIM[Br], highlighting cholesterol's strong stabilizing effect. For HMIM[Br] at 300 K and high concentration (50 mol%), a minor secondary peak appears around ~1100 nm-approximately 20 times larger than the original vesicle size-suggesting limited aggregation (Fig. 9a). However, this aggregation is significantly reduced compared to cholesterol-free DPPC ULVs[24], as supported by SANS data showing weaker bilayer-bilayer interactions and lower volume fractions. In the case of DMIM[Br], the size distribution remains largely unchanged even at 330 K, confirming minimal impact on vesicle morphology in the presence of cholesterol (Fig. 9b). In contrast, pure DPPC ULVs exhibit strong sensitivity to IL-induced aggregation, particularly in the presence of HMIM[Br]. At 300 K, DPPC ULVs display a monomodal size distribution centered around ~58 nm. Upon addition of 25 mol% HMIM[Br], the distribution becomes bimodal, with a secondary peak at much larger sizes-about 20 times greater-indicating vesicle aggregation (Fig. 9c)[24]. The extent of aggregation increases further at 50 mol%. At 330 K, aggregation persists but is attenuated, as reflected in a less prominent secondary peak (Fig. 9c), consistent with SANS results showing weaker inter-vesicle interactions at elevated temperature. In contrast, DMIM[Br] does not induce aggregation at 300 K (Fig. 9d)[24]. At 330 K, a slight broadening of the size distribution and a modest shift (~1.4 times increase) are observed (Fig. 9d), suggesting subtle effects on vesicle structure without clear aggregation. These observations align well with the SANS data, further confirming that the longer alkyl chain in DMIM[Br] leads to a fundamentally different and milder mode of interaction compared to HMIM[Br].

These findings collectively indicate that cholesterol acts as a molecular stabilizer, enhancing membrane resilience against IL-induced perturbations through a combination of physical (increased rigidity and reduced free volume) and chemical (hydrogen bonding and acyl chain interactions) mechanisms.

**Role of cholesterol on membrane permeabilization in the presence of ILs**

To investigate the role of cholesterol on the IL-induced membrane permeability, dye leakage measurements were carried out on ULVs formed by DPPC in presence of cholesterol with varied concentrations of HMIM[Br] and DMIM[Br] at 320 K. Observed time-dependent dye-leakages as measured using fluorescence assay are shown in Fig. 10. For direct comparison, time dependent dye-leakage measured for pure DPPC with varying concentrations of ILs at 320 K are also shown[24]. The dye leakage data due to the presence of triton, serving as a positive control, is also included in Fig. 10. It is evident that the impact of ILs on membrane permeability is significantly modulated by the presence of cholesterol in DPPC ULVs. As illustrated in Fig. 10(a & b), DPPC-cholesterol ULVs exhibit markedly different dye leakage behavior upon incorporation of HMIM[Br] and DMIM[Br]. The addition of HMIM[Br] does not induce any significant dye release within the measured concentration range, with the maximum leakage observed being only ~3% (Fig. 10a). In contrast, DMIM[Br] incorporation leads to a pronounced increase in leakage from DPPC-cholesterol ULVs, reaching up to ~99% (Fig. 10b). This stark contrast with HMIM[Br] highlights the critical role of alkyl chain length in modulating IL-membrane interactions and leakage behavior.

In comparison, pure DPPC ULVs display distinct leakage in response to these ILs.[24] It is evident that in the case of HMIM[Br], the maximum dye leakage from pure DPPC ULVs reaches approximately 13% (Fig. 10c), which is significantly higher than the ~3% observed for DPPC ULVs containing cholesterol. However, for the longer-chain IL, DMIM[Br], the leakage from pure DPPC ULVs (Fig. 10d) is comparatively lower, and the time required to reach maximum leakage is longer than that observed for DPPC-cholesterol ULVs. This striking contrast highlights that cholesterol's protective effect is not uniformly effective across all ILs; rather, it is strongly influenced by the specific molecular structure of the IL.

Given the substantial impact of DMIM[Br] on the permeability of both ULVs, we conducted a detailed analysis of the time-dependent dye leakage kinetics for a direct comparison between DPPC membranes with and without cholesterol. The % dye leakage data were fitted using the following sigmoidal equation[21]

$$\% \ dye\ leakage = \frac{a_0}{1+e^{-K_r(t-t_c)}} \qquad (6)$$

where $a_0$ is the maximum leakage, $K_r$ is the rate constant and $t_c$ is the leakage time. The above equation was used to describe the dye leakage data for longer chain IL DMIM[Br]. It is found this model describes the data well for both vesicle systems composed of DPPC and DPPC with cholesterol. Solid lines in Fig. 10 (b) & (d) show the fits of the dye leakage data for both ULVs. The rate constant ($K_r$) signifies the rate at which dye is released from ULVs, while the leakage time ($t_c$) represents the time when dye leakage reduces to half of the maximum leakage ($a_0$). A comprehensive comparison of the effect of various concentrations of DMIM[Br] on both ULVs with and without cholesterol, featuring $a_0$, $K_r$, and $t_c$, is presented in Fig. 11 (a), (b), and (c), respectively.

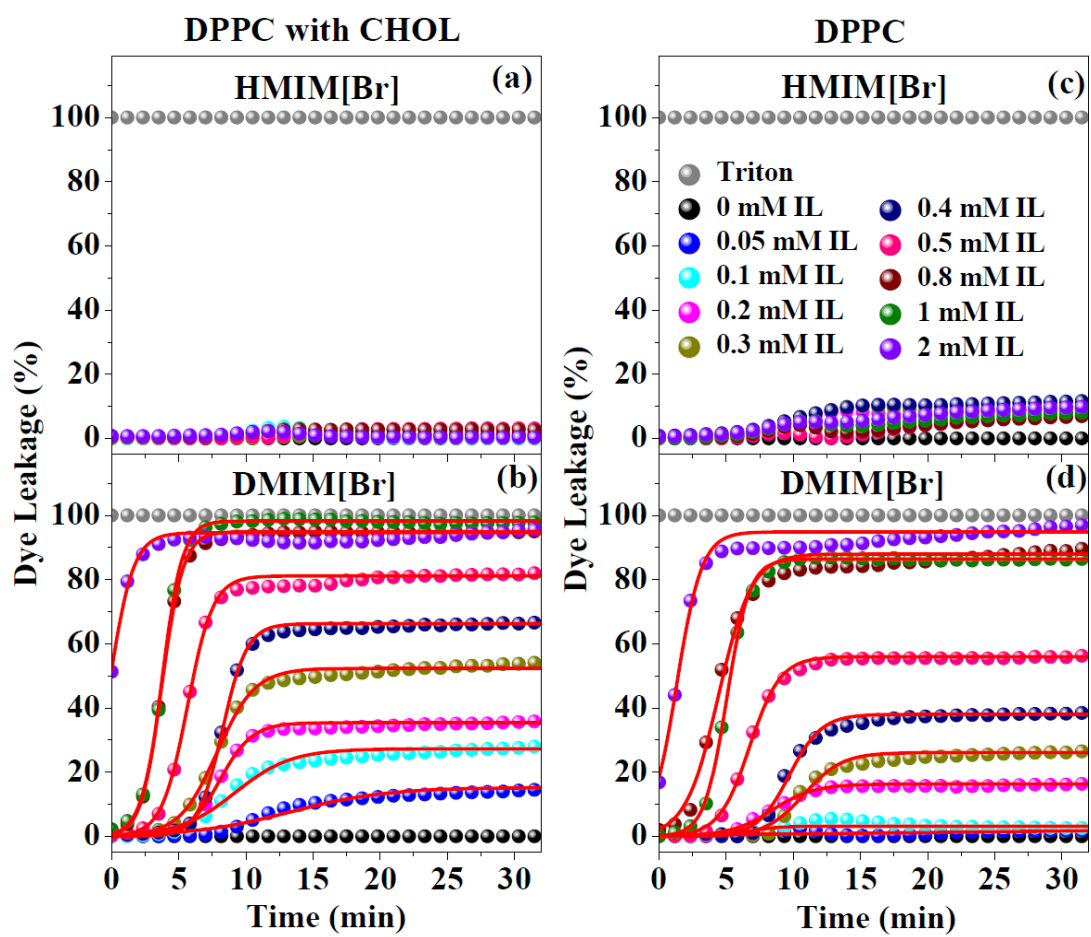

**FIGURE 10:** Time course of calcein dye leakage due to presence of (a) HMIM[Br] and (b) DMIM[Br] in the ULVs composed of DPPC with cholesterol. For direct comparison, dye leakage data for pure DPPC in presence of (c) HMIM[Br] and (d) DMIM[Br] are taken from ref[24]. All the measurements were performed in the fluid phase of membrane. Solid red lines represent sigmoidal fittings.

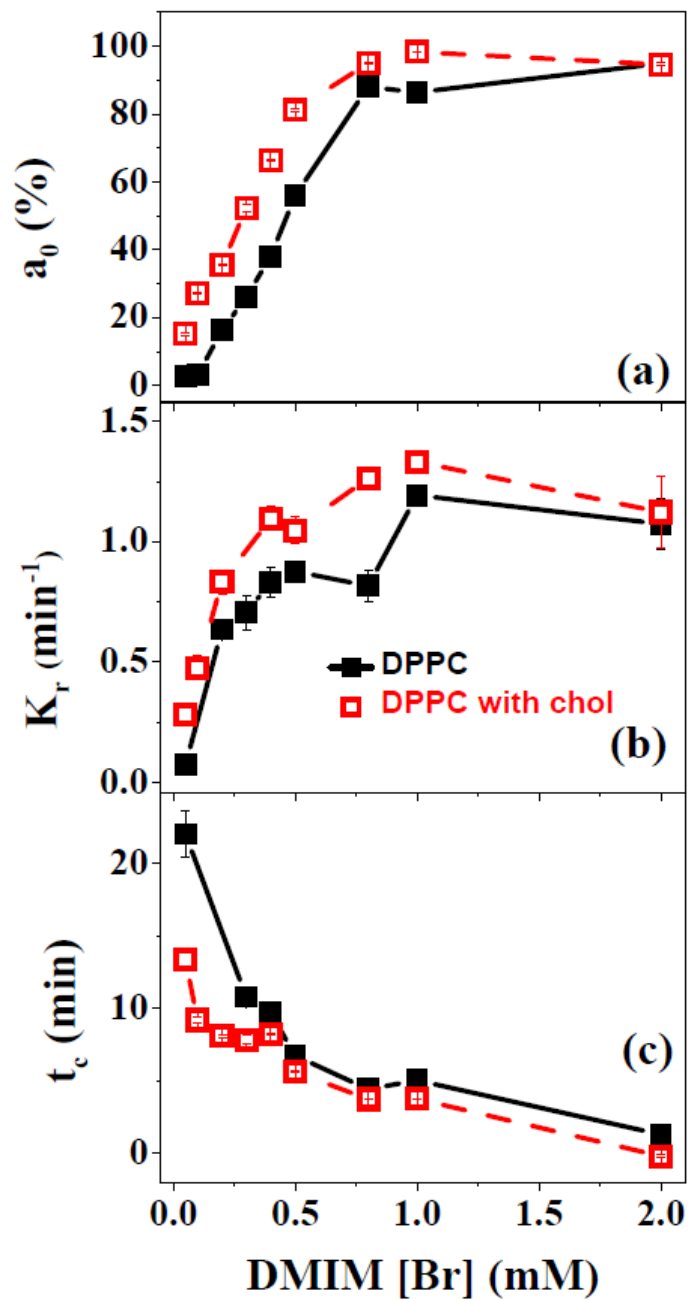

**FIGURE 11:** (a) Maximum dye leakage ($a_0$), (b) rate constant ($K_r$), and (c) half time ($t_c$) of dye leakage from the fluid phase of DPPC ULVs (filled symbol) and DPPC with cholesterol ULVs (open symbol) in response to increasing concentration of DMIM[Br] IL.

At a low concentration of DMIM[Br] (ca 0.05 mM) in DPPC ULVs, the results indicate an insignificant $a_0$, signifying negligible dye release from the ULVs. Simultaneously, a lower $K_r$ and a large $t_c$ value suggest a slow dye release process with a prolonged time required for half of the dye to decay. With an increase in DMIM[Br] concentration, both $a_0$ and $K_r$ show an upward trend, accompanied by a progressive decrease in $t_c$, indicating a shorter time required for leakage. At 2 mM DMIM[Br], $a_0$ reaches approximately 97%. Collectively, these parameters suggest that an increasing concentration of the ILs enhances the permeability of DPPC ULVs.

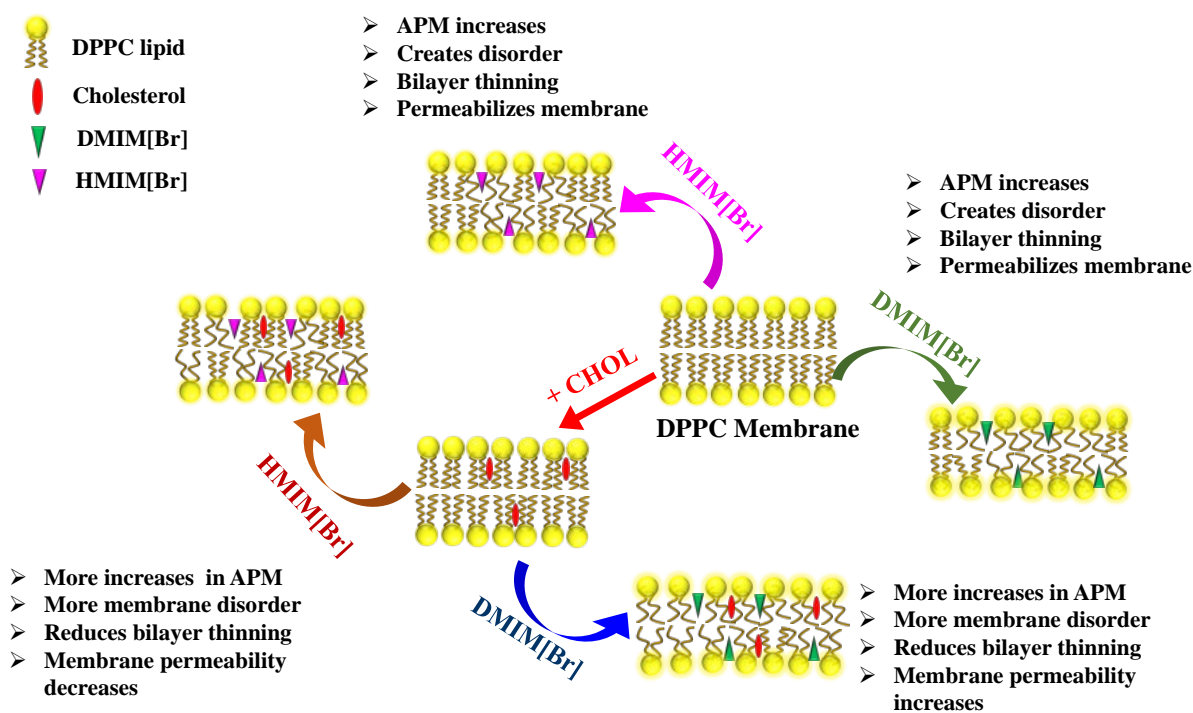

**FIGURE 12**. Schematic representation illustrating the role of cholesterol in modulating IL-membrane interactions. In the absence of cholesterol, ILs readily insert and disrupt the lipid bilayer, leading to increased membrane disorder, area per molecule (APM) and bilayer thinning. In contrast, cholesterol-rich membranes exhibit enhanced packing and rigidity, thereby reducing IL penetration and stabilizing the membrane structure. ILs induced changes in APM, membrane disorder, bilayer thickness and membrane permeability altered due to cholesterol highlighting presence of cholesterol's modulates IL-membrane interaction

In Fig. 11, all parameters ($a_0$, $K_r$, and $t_c$) obtained for pure DPPC and DPPC with cholesterol ULVs at varying concentrations of DMIM[Br] are presented. The results reveal not only higher maximum leakage but also a faster leakage rate in the case of the DPPC-cholesterol membrane. This suggests that in the presence of cholesterol, the membrane is more prone to leakage due to DMIM[Br] compared to the neat DPPC membrane. This finding is counterintuitive and sharply contrasts with previous studies[12-13], which reported that the disturbing effects of ILs are inhibited in the presence of cholesterol. It's worth noting that the previous studies were conducted on either sphingomyelin[12] or unsaturated PC/PG lipids[21], and different ILs were used. Therefore, the interaction of the IL with the lipid membrane in the presence of cholesterol involves a complex interplay of various parameters, including the molecular structure of the lipid and the ILs. Despite the weaker binding of DMIM[Br] to DPPC-cholesterol membranes compared to pure DPPC, the DMIM[Br] significantly enhances membrane permeabilization in the presence of cholesterol. This observation suggests that binding strength alone is not the sole determinant of permeabilization. The increased susceptibility of the DPPC-cholesterol membrane to permeabilization could be attributed to cholesterol's role in ordering the membrane. This increased order, while stabilizing the bilayer, makes the membrane more vulnerable to stress or defects introduced by the IL. The ordered phase constrains the membrane's ability to recover from disruptions, amplifying the permeabilizing effect of DMIM[Br].

## 5. CONCLUSIONS

This study elucidates the pivotal role of cholesterol in modulating the interaction between ionic liquids (ILs) and lipid membranes, offering a nuanced understanding through complementary analyses of monolayers and unilamellar vesicles (ULVs). The schematic representation of role of cholesterol on the IL-membrane interaction is shown in Fig. 12. Surface pressure-area isotherm measurements demonstrate that IL incorporation into lipid monolayers increases the area per molecule (APM) in a concentration-dependent manner. The effect is particularly pronounced for longer-chain ILs, such as DMIM[Br], indicating their more substantial impact on lipid organization. Cholesterol introduces an additional layer of complexity: it compacts the lipid arrangement, reducing the baseline APM, but its presence amplifies the IL-induced APM increase, especially in DPPC/cholesterol membranes compared to pure DPPC membranes. This arrangement highlights role of cholesterol in enhancing the membrane's responsiveness to ILs, with variations depending on the IL's structure.

Isothermal titration calorimetry (ITC) data reveal that cholesterol reduces both the association constant and the Gibbs free energy of IL binding, indicating weakened IL-membrane binding in cholesterol-containing systems. However, complementary rheological, FTIR, and dye-leakage analyses suggest that cholesterol significantly alters the interaction dynamics, with the effects dictated by the IL's alkyl chain length. For shorter-chain ILs like HMIM[Br], electrostatic interactions dominate, and cholesterol mitigates membrane perturbation, as evidenced by smaller blueshift, reduced HWHM changes in $CH_2$ peaks, and diminished dye leakage. In contrast, for longer-chain ILs like DMIM[Br], hydrophobic interactions are more prominent, and cholesterol amplifies their disruptive effects, reflected in greater blueshifts, larger HWHM changes, and enhanced dye leakage.

Despite cholesterol's dampening of IL binding strength-particularly for DMIM[Br]-it paradoxically enhances membrane permeabilization. This underscores that membrane disruption is not solely dependent on binding strength but is also influenced by the IL's molecular structure and the membrane's organization. FTIR results show that cholesterol promotes a more ordered membrane phase, which, while stabilizing the lipid structure, makes the membrane more susceptible to disruption by ILs like DMIM[Br]. This ordered phase compromises the membrane's resilience, increasing its permeability under IL-induced stress.

This study highlights that longer-chain molecules with hydrophobic interactions significantly disrupt cholesterol-enriched membranes. Due to their well-ordered structure, these membranes are more susceptible to rupture when long-chain molecules insert themselves into the membrane. Overall, the interplay between IL structure, lipid composition, and cholesterol's ordering effects governs the membrane's integrity and permeability. The study provides critical insights into the multifaceted mechanisms underlying IL-membrane interactions in the presence of cholesterol, emphasizing the need to consider both molecular binding and structural dynamics. These findings have far-reaching implications for designing IL-based antimicrobial agents, highlighting the significance of lipid composition in modulating IL efficacy and paving the way for more targeted and effective strategies.

## 6. SUPPORTING INFORMATION

Theoretical framework for analysis of DLS, SANS and ITC data, DSC and FTIR data for DPPC and DPPC with cholesterol in absence and presence of ILs, Temperature dependent peak position and HWHM of $\nu_s(CH_2)$ for DPPC with and without cholesterol in absence and presence of ILs.

## 7. ACKNOWLEDGMENTS

The authors sincerely thank Dr. H. Bhatt, BARC for his help with FTIR measurements. The authors extend their heartfelt gratitude to Dr. V. K. Aswal (Head, SSPD, BARC) for his unwavering support and encouragement.

# 8. REFERENCES

1. Hayes, R.; Warr, G. G.; Atkin, R., Structure and Nanostructure in Ionic Liquids. *Chemical Reviews* **2015**, *115*, 6357-6426.
2. Tan, C., et al., Recent Advances in Ultrathin Two-Dimensional Nanomaterials. *Chemical Reviews* **2017**, *117*, 6225-6331.
3. Nikfarjam, N., et al., Antimicrobial Ionic Liquid-Based Materials for Biomedical Applications. *Advanced Functional Materials* **2021**, *31*, 2104148.
4. Welton, T., Room-Temperature Ionic Liquids. Solvents for Synthesis and Catalysis. *Chemical Reviews* **1999**, *99*, 2071-2084.
5. Sharma, V. K.; Gupta, J.; Mitra, J. B.; Srinivasan, H.; Sakai, V. G.; Ghosh, S. K.; Mitra, S., The Physics of Antimicrobial Activity of Ionic Liquids. *The Journal of Physical Chemistry Letters* **2024**, *15*, 7075-7083.
6. Egorova, K. S.; Kibardin, A. V.; Posvyatenko, A. V.; Ananikov, V. P., Mechanisms of Biological Effects of Ionic Liquids: From Single Cells to Multicellular Organisms. *Chemical Reviews* **2024**, *124*, 4679-4733.
7. Bakshi, K.; Mitra, S.; Sharma, V. K.; Jayadev, M. S. K.; Sakai, V. G.; Mukhopadhyay, R.; Gupta, A.; Ghosh, S. K., Imidazolium-Based Ionic Liquids Cause Mammalian Cell Death Due to Modulated Structures and Dynamics of Cellular Membrane. *Biochimica et Biophysica Acta (BBA) - Biomembranes* **2020**, *1862*, 183103.
8. Jing, B.; Lan, N.; Qiu, J.; Zhu, Y., Interaction of Ionic Liquids with a Lipid Bilayer: A Biophysical Study of Ionic Liquid Cytotoxicity. *The journal of physical chemistry. B* **2016**, *120*, 2781-9.
9. Jamieson, G. A.; Robinson, D. M., *Mammalian Cell Membranes: Volume 2: The Diversity of Membranes*; Elsevier, 2014.
10. Mouritsen, O. G.; Zuckermann, M. J., What's So Special About Cholesterol? *Lipids* **2004**, *39*, 1101-1113.
11. Genova, J.; Bivas, I.; Marinov, R., Cholesterol Influence on the Bending Elasticity of Lipid Membranes. *Colloids and Surfaces A: Physicochemical and Engineering Aspects* **2014**, *460*, 79-82.
12. Hitaishi, P.; Raval, M.; Seth, A.; Kumar, S.; Mithu, V. S.; Sharma, V. K.; Ghosh, S. K., Cholesterol-Controlled Interaction of Ionic Liquids with Model Cellular Membranes. *Langmuir* **2023**, *39*, 9396-9405.
13. Hao, X.-L.; Cao, B.; Dai, D.; Wu, F.-G.; Yu, Z.-W., Cholesterol Protects the Liquid-Ordered Phase of Raft Model Membranes from the Destructive Effect of Ionic Liquids. *The Journal of Physical Chemistry Letters* **2022**, *13*, 7386-7391.
14. Jurak, M., Thermodynamic Aspects of Cholesterol Effect on Properties of Phospholipid Monolayers: Langmuir and Langmuir–Blodgett Monolayer Study. *The Journal of Physical Chemistry B* **2013**, *117*, 3496-3502.
15. Almeida, P. F. F.; Pokorny, A.; Hinderliter, A., Thermodynamics of Membrane Domains. *Biochimica et Biophysica Acta (BBA) - Biomembranes* **2005**, *1720*, 1-13.
16. Huang, J.; Feigenson, G. W., A Microscopic Interaction Model of Maximum Solubility of Cholesterol in Lipid Bilayers. *Biophysical journal* **1999**, *76*, 2142-57.
17. Sharma, V. K.; Mamontov, E.; Anunciado, D. B.; O'Neill, H.; Urban, V. S., Effect of Antimicrobial Peptide on the Dynamics of Phosphocholine Membrane: Role of Cholesterol and Physical State of Bilayer. *Soft Matter* **2015**, *11*, 6755-6767.
18. Wang, Y.; Gkeka, P.; Fuchs, J. E.; Liedl, K. R.; Cournia, Z., Dppc-Cholesterol Phase Diagram Using Coarse-Grained Molecular Dynamics Simulations. *Biochimica et biophysica acta* **2016**, *1858*, 2846-2857.

19. Krause, M. R.; Regen, S. L., The Structural Role of Cholesterol in Cell Membranes: From Condensed Bilayers to Lipid Rafts. *Accounts of Chemical Research* **2014**, *47*, 3512-3521.
20. Russo, G.; Witos, J.; Rantamäki, A. H.; Wiedmer, S. K., Cholesterol Affects the Interaction between an Ionic Liquid and Phospholipid Vesicles. A Study by Differential Scanning Calorimetry and Nanoplasmonic Sensing. *Biochimica et Biophysica Acta (BBA) - Biomembranes* **2017**, *1859*, 2361-2372.
21. Kumar, S.; Kaur, N.; Hitaishi, P.; Ghosh, S. K.; Mithu, V. S.; Scheidt, H. A., Role of Cholesterol in Interaction of Ionic Liquids with Model Lipid Membranes and Associated Permeability. *The Journal of Physical Chemistry B* **2024**, *128*, 5407-5418.
22. Sharma, V. K.; Gupta, J.; Srinivasan, H.; Bhatt, H.; García Sakai, V.; Mitra, S., Curcumin Accelerates the Lateral Motion of Dppc Membranes. *Langmuir : the ACS journal of surfaces and colloids* **2022**, *38*, 9649-9659.
23. Sharma, V. K.; Gupta, J.; Srinivasan, H.; Hitaishi, P.; Ghosh, S. K.; Mitra, S., Quantifying Ionic Liquid Affinity and Its Effect on Phospholipid Membrane Structure and Dynamics. *Langmuir : the ACS journal of surfaces and colloids* **2025**, *41*, 11547-11562.
24. Gupta, J.; Sharma, V. K.; Hitaishi, P.; Srinivasan, H.; Kumar, S.; Ghosh, S. K.; Mitra, S., Structural Reorganizations and Nanodomain Emergence in Lipid Membranes Driven by Ionic Liquids. *Langmuir : the ACS journal of surfaces and colloids* **2025**, *41*, 79-90.
25. Hitaishi, P.; Seth, A.; Mitra, S.; Ghosh, S. K., Thermodynamics and in-Plane Viscoelasticity of Anionic Phospholipid Membranes Modulated by an Ionic Liquid. *Pharmaceutical Research* **2022**, *39*, 2447-2458.
26. Miller, R.; Ferri, J. K.; Javadi, A.; Krägel, J.; Mucic, N.; Wüstneck, R., Rheology of Interfacial Layers. *Colloid and Polymer Science* **2010**, *288*, 937-950.
27. Macosko, C. W., Rheology: Principles, Measurements, and Applications. *Wiley* **1994**.
28. Mendoza, A. J.; Guzmán, E.; Martínez-Pedrero, F.; Ritacco, H.; Rubio, R. G.; Ortega, F.; Starov, V. M.; Miller, R., Particle Laden Fluid Interfaces: Dynamics and Interfacial Rheology. *Advances in Colloid and Interface Science* **2014**, *206*, 303-319.
29. Miller, R.; Liggieri, L., Interfacial Rheology. **2009**.
30. Aswal, V. K.; Goyal, P. S., Small-Angle Neutron Scattering Diffractometer at Dhruva Reactor. *Current Science (Bangalore)* **2000**, *79*, 947-953.
31. Mitra, S.; Sharma, V. K.; Mitra, J. B.; Chowdhury, S.; Mukhopadhyay, M. K.; Mukhopadhyay, R.; Ghosh, S. K., Thermodynamics and Structure of Model Bio-Membrane of Liver Lipids in Presence of Imidazolium-Based Ionic Liquids. *Biochimica et Biophysica Acta (BBA) - Biomembranes* **2021**, *1863*, 183589.
32. Hitaishi, P.; Mandal, P.; Ghosh, S. K., Partitioning of a Hybrid Lipid in Domains of Saturated and Unsaturated Lipids in a Model Cellular Membrane. *ACS Omega* **2021**, *6*, 34546-34554.
33. Mitra, S.; Bhattacharya, G.; Ghosh, S. K., Quantifying the Structural Effects of Ionic Liquids on Model Cellular Membrane. **2019**.
34. Mitra, S.; Ray, D.; Bhattacharya, G.; Gupta, R.; Sen, D.; Aswal, V. K.; Ghosh, S. K., Probing the Effect of a Room Temperature Ionic Liquid on Phospholipid Membranes in Multilamellar Vesicles. *European Biophysics Journal* **2019**, *48*, 119-129.
35. Bhattacharya, G.; Mitra, S.; Mandal, P.; Dutta, S.; Giri, R. P.; Ghosh, S. K., Thermodynamics of Interaction of Ionic Liquids with Lipid Monolayer. *Biophysical Reviews* **2018**, *10*, 709-719.
36. Kaur, N.; Fischer, M.; Hitaishi, P.; Kumar, S.; Sharma, V. K.; Ghosh, S. K.; Gahlay, G. K.; Scheidt, H. A.; Mithu, V. S., How 1,N-Bis(3-Alkylimidazolium-1-Yl) Alkane Interacts with the Phospholipid Membrane and Impacts the Toxicity of Dicationic Ionic Liquids. *Langmuir : the ACS journal of surfaces and colloids* **2022**, *38*, 13803-13813.

37. Mitra, S.; Das, R.; Singh, A.; Mukhopadhyay, M. K.; Roy, G.; Ghosh, S. K., Surface Activities of a Lipid Analogue Room-Temperature Ionic Liquid and Its Effects on Phospholipid Membrane. *Langmuir : the ACS journal of surfaces and colloids* **2020**, *36*, 328-339.
38. Mendonça, C. M. N., et al., Understanding the Interactions of Imidazolium-Based Ionic Liquids with Cell Membrane Models. *Physical Chemistry Chemical Physics* **2018**, *20*, 29764-29777.
39. Geraldo, V. P. N.; Pavinatto, F. J.; Nobre, T. M.; Caseli, L.; Oliveira, O. N., Langmuir Films Containing Ibuprofen and Phospholipids. *Chemical Physics Letters* **2013**, *559*, 99-106.
40. Regen, S. L., Cholesterol's Condensing Effect: Unpacking a Century-Old Mystery. *JACS Au* **2022**, *2*, 84-91.
41. Bhattacharya, G.; Giri, R. P.; Dubey, A.; Mitra, S.; Priyadarshini, R.; Gupta, A.; Mukhopadhyay, M. K.; Ghosh, S. K., Structural Changes in Cellular Membranes Induced by Ionic Liquids: From Model to Bacterial Membranes. *Chemistry and Physics of Lipids* **2018**, *215*, 1-10.
42. Singh, P.; Sharma, V. K.; Singha, S.; García Sakai, V.; Mukhopadhyay, R.; Das, R.; Pal, S. K., Unraveling the Role of Monoolein in Fluidity and Dynamical Response of a Mixed Cationic Lipid Bilayer. *Langmuir : the ACS journal of surfaces and colloids* **2019**, *35*, 4682-4692.
43. Gupta, J.; Sharma, V. K.; Srinivasan, H.; Bhatt, H.; Kumar, S.; Sarter, M.; Sakai, V. G.; Mitra, S., Microscopic Diffusion in Cationic Vesicles across Different Phases. *Physical Review Materials* **2022**, *6*, 075602.
44. Severcan, F.; Baykal, U.; Süzer, S., Ftir Studies of Vitamin E-Cholesterol-Dppc Membrane Interactions in Ch(2) Region. *Analytical and bioanalytical chemistry* **1996**, *355*, 415-7.
45. Pöhnl, M.; Trollmann, M. F. W.; Böckmann, R. A., Nonuniversal Impact of Cholesterol on Membranes Mobility, Curvature Sensing and Elasticity. *Nature Communications* **2023**, *14*, 8038.
46. Wu, F.-G.; Wang, N.-N.; Yu, Z.-W., Nonsynchronous Change in the Head and Tail of Dioctadecyldimethylammonium Bromide Molecules During the Liquid Crystalline to Coagel Phase Transformation Process. *Langmuir : the ACS journal of surfaces and colloids* **2009**, *25*, 13394-13401.
47. Sharma, V. K.; Ghosh, S. K.; Mandal, P.; Yamada, T.; Shibata, K.; Mitra, S.; Mukhopadhyay, R., Effects of Ionic Liquids on the Nanoscopic Dynamics and Phase Behaviour of a Phosphatidylcholine Membrane. *Soft Matter* **2017**, *13*, 8969-8979.
48. Pencer, J.; Nieh, M.-P.; Harroun, T. A.; Krueger, S.; Adams, C.; Katsaras, J., Bilayer Thickness and Thermal Response of Dimyristoylphosphatidylcholine Unilamellar Vesicles Containing Cholesterol, Ergosterol and Lanosterol: A Small-Angle Neutron Scattering Study. *Biochimica et Biophysica Acta (BBA) - Biomembranes* **2005**, *1720*, 84-91.
49. Róg, T.; Pasenkiewicz-Gierula, M.; Vattulainen, I.; Karttunen, M., Ordering Effects of Cholesterol and Its Analogues. *Biochimica et Biophysica Acta (BBA) - Biomembranes* **2009**, *1788*, 97-121.
50. Provencher, S., Contin: A General Purpose Constrained Regularization Program for Inverting Noisy Linear Algebraic and Integral Equations. *Comput. Phys. Commun.* **1982** *27*, 229-242.